\documentclass[vancouver,Times1COL]{WileyNJDv5}

\usepackage{float}
\articletype{Article Type}%

\startpage{1}

\begin{document}

\title{A Time-Series Model for Areal Data Using Area-Specific Gaussian Processes with Spatially Correlated Hyperparameters}

\author[1,2]{Alejandro Rozo Posada}%[https://orcid.org/0000-0001-5483-2222]
\author[2]{Oswaldo Gressani}%[https://orcid.org/0000-0003-4152-6159]
\author[2]{Christel Faes}%[https://orcid.org/0000-0002-1878-9869]
\author[3]{James Colborn}
\author[4]{Baltazar Candrinho}
\author[5]{Emanuele Giorgi$^\dagger$}%[https://orcid.org/0000-0003-0640-181X]
\author[1,2]{Thomas Neyens$^\dagger$}%[https://orcid.org/0000-0003-2364-7555]
\authormark{Rozo Posada \textsc{et al.}}
\titlemark{A Time-Series Model for Areal Data Using Area-Specific Gaussian Processes with Spatially Correlated Hyperparameters}

\address[1]{\orgdiv{Leuven Biostatistics and Statistical Bioinformatics Centre (L-Biostat),}\orgname{KU Leuven, }%
\orgaddress{\state{Leuven, }\country{Belgium}}}

\address[2]{\orgdiv{Data Science Institute (DSI), }\orgname{Hasselt University, }%
\orgaddress{\state{Hasselt, }\country{Belgium}}}

\address[3]{\orgdiv{Clinton Health Access Initiative, }%
\orgaddress{\state{Maputo, }\country{Mozambique}}}

\address[4]{\orgdiv{National Malaria Control Program, }\orgname{Ministry of Health, }%
\orgaddress{\state{Maputo, }\country{Mozambique}}}

\address[5]{\orgdiv{Department of Applied Health Sciences, }\orgname{University of Birmingham, }%
\orgaddress{\state{Birmingham, }\country{England}}}

\corres{Alejandro Rozo Posada, Leuven Biostatistics and Statistical Bioinformatics Centre (L-Biostat), KU Leuven, Leuven, Belgium. \email{josealejandro.rozoposada@kuleuven.be}}

%\fundingInfo{Text}
%\JELinfo{ejlje}

\abstract[Abstract]{In many applied settings, areal data are observed repeatedly over long time periods, a structure that frequently arises in infectious disease surveillance as well as in environmental and demographic monitoring. Accurate characterization of local temporal dynamics and uncertainty is essential for monitoring disease trends and supporting public health decision-making. Traditional spatio-temporal models for such data typically represent spatial, temporal, and space-time interaction components through structured random effects acting on the latent outcome process.
We propose a Bayesian spatio-temporal hierarchical modeling framework in which temporal dynamics are modeled through area-specific Gaussian processes and spatial dependence is introduced by modeling the Gaussian-process covariance hyperparameters as spatially correlated. This formulation allows neighboring regions to share information about the characteristics of their temporal dependence while preserving area-specific temporal trajectories, providing an alternative representation of spatio-temporal dependence.
Inference is conducted using Markov chain Monte Carlo methods. The methodology is illustrated using monthly malaria incidence data from three Mozambican provinces. Predictive performance is evaluated using the Root Mean Squared Error, Continuous Ranked Probability Score, empirical coverage probability, and credible interval width. Compared with established spatio-temporal models, the proposed framework achieves competitive predictive accuracy while yielding better-calibrated predictive uncertainty. Incorporating spatial dependence into the temporal hyperparameters improves predictive interval calibration relative to an equivalent model with independent temporal processes, demonstrating that this formulation provides a competitive alternative to conventional outcome-level spatial smoothing for the analysis of spatio-temporal areal data.
}


%\jnlcitation{\cname{%
%\author{Taylor M.},
%\author{Lauritzen P},
%\author{Erath C}, and
%\author{Mittal R}}.
%\ctitle{On simplifying ‘incremental remap’-based transport schemes.} \cjournal{\it J Comput Phys.} \cvol{2021;00(00):1--18}.}

\maketitle
\renewcommand\thefootnote{\fnsymbol{footnote}}
\setcounter{footnote}{0}
\footnotetext[2]{$^\dagger$Senior authorship was shared equally by these authors.}

%\renewcommand\thefootnote{}
%\footnotetext{\textbf{Abbreviations:} ANA, anti-nuclear antibodies; APC, antigen-presenting cells; IRF, interferon regulatory factor.}

%\renewcommand\thefootnote{\fnsymbol{footnote}}
%\setcounter{footnote}{1}

\section{Introduction}\label{Section.Intro}

In public health and epidemiological research, data are often collected at multiple geographic locations and over time. These observations are often aggregated within non-overlapping administrative areas such as districts, counties, or municipalities.  These spatially aggregated observations are commonly referred to as areal data. Temporal measurements of these data can occur at different resolutions, from days to months or years, resulting in complex spatio-temporal structures.  For instance, in disease mapping, the incidence of an infectious disease, recorded across geographic regions and over time, can reveal patterns of transmission and outbreak dynamics \cite{dismap1Thai, dismap2Aust, dismapViet}. Similarly,  in environmental health studies, longitudinal areal data are used to assess the health impact of environmental stressors such as heatwaves \cite{heatwaves1, heatwaves2}.  Demographic and socioeconomic indicators can be tracked to understand geographical shifts in population trends \cite{socialind1, socialind2}.  Across these applications, a fundamental statistical challenge is accounting simultaneously for spatial dependence, where nearby areas may exhibit correlated outcomes, and temporal dependence, including both short- and long-term trends. These challenges are more complicated when data are sparse or irregularly observed.  Addressing them requires advanced modeling approaches that can appropriately capture spatio-temporal dependencies while providing reliable inference. \\

A key difficulty in analyzing such data arises from the common assumption of conditional independence among observations given explanatory variables. This assumption is often violated due to spatial, temporal, or spatio-temporal correlation, which, when not taken into account, may lead to biased parameter estimates, underestimated uncertainty, and invalid inference \cite{lee2018spatio, rushworth2014spatio, gomezbayesian}. To overcome these limitations, a variety of spatio-temporal models have been developed to explicitly account for both spatial and temporal dependencies. For a comprehensive overview of these models, the reader is referred to Wang et al.\cite{review1_st} and Nazia et al.\cite{review2_st}. These models exploit the correlation structure in the data to improve efficiency and predictive accuracy while preventing overestimation of the effective sample size that would result from assuming independence across space and time. By borrowing information across regions and time periods, spatio-temporal (ST) models produce more stable and reliable estimates than those derived from models that treat each dimension in isolation. Examples of such simpler approaches include area-specific (S)ARIMA or Gaussian-process time-series models \cite{box2015time,diggle2025time}, or cross-sectional spatial models parametrized with Besag-York-Mollié (BYM) \cite{besag1991bayesian} or Leroux \cite{leroux2000estimation} random effects. \\

Hierarchical Bayesian ST models are typically specified with latent random effects that capture unobserved patterns across space, time, and their interaction. These effects are often parameterized as area-specific, time-specific, or region-by-time-specific random terms. A wide range of prior distributions has been proposed for these components. Spatial effects are commonly modeled using (adaptive) conditional autoregressive (CAR) priors or BYM structures. Temporal effects are often modeled using autoregressive (AR) processes or random walks (RWs). Interaction terms for spatio-temporal components can be specified in various ways, including additive \cite{knorr2000bayesian, ugarte2012gender, macnab2023adaptive}, multiplicative \cite{bernardinelli1995bayesian}, or separable structures \cite{rushworth2014spatio, napier2016model}. \\

Despite their flexibility, many hierarchical ST models exhibit notable limitations. A central drawback is their difficulty in simultaneously representing localized temporal trends while maintaining appropriate smoothing across regions and time periods \cite{rushworth2017adaptive}. This issue often arises in standard models that use global or partially pooled temporal parameters, effectively imposing a common temporal structure across all regions. Such assumptions are common when ST models are conceived as natural extensions of spatial modeling frameworks; however, they can be restrictive in practice. Regions often exhibit heterogeneous temporal dynamics driven by local epidemiological, environmental, or demographic factors. Failure to accommodate this heterogeneity may result in biased parameter estimation, underestimated uncertainty, and diminished predictive performance. An additional limitation is the treatment of time as discrete with short-range dependence, typically specified through first- or second-order autoregressive structures. In these formulations, temporal correlation is confined to a small number of preceding time points. While computationally convenient, such structures may be inadequate for long time series, where dependence can extend and exhibit varying smoothness or persistence. Furthermore, when temporal dependence is strong and persistent, inference on spatial variation may become increasingly sensitive to the specification of the temporal component, particularly when temporal dynamics differ across regions. Models that impose a common temporal structure may therefore struggle to disentangle local temporal variation from residual spatial patterns, potentially affecting both estimation and uncertainty quantification.\\

Addressing these limitations is essential not only in disease mapping and environmental applications but in any context requiring reliable region-level inference, including public health surveillance, targeted intervention planning, and small area estimation. Developing modeling frameworks that allow area-specific temporal trajectories while coherently borrowing strength across space and time can yield more accurate inference, improved forecasting, and more interpretable estimates for decision-making.\\

To tackle these challenges, we propose a novel approach that represents spatio-temporal data as a collection of area-specific temporal processes whose parameters vary smoothly across space. Specifically, we model each region's time series using a Gaussian process (GP), while introducing spatial structure through a hierarchical prior on the GP hyperparameters. Gaussian processes provide a flexible framework for modeling temporal dependence because they characterize correlation through a covariance function rather than a fixed autoregressive structure, allowing key features of the temporal dynamics, such as their variability and range, to be learned from the data. This flexibility is particularly attractive in long time series, where dependence may extend across multiple time scales and vary substantially across regions. Spatial information is incorporated by imposing a conditional autoregressive (CAR) prior on the variance parameters of the area-specific GPs and by linking the temporal range to the level of temporal variability. Although the latent GP realizations remain conditionally independent across regions, spatial dependence arises marginally through the shared hierarchical structure, enabling partial pooling of information across neighboring areas. By regularizing area-specific temporal processes through spatially correlated hyperparameters, the proposed framework encourages neighboring regions to exhibit similar temporal characteristics while preserving regional interpretability and improving estimation stability in settings with limited data. \\

The remainder of this paper is organized as follows. Section 2 introduces the case study we used to develop and evaluate our method. Sections 3 and 4 present standard spatio-temporal models and our proposed model, respectively, detailing their formulations and estimation approaches, and introducing our comparison procedure. Section 5 presents our findings, and Section 6 offers an in-depth discussion of the results. Finally, section 7 provides a conclusion.

\section{Motivating example: Malaria incidence in Mozambique} \label{Section.examples}
The motivating example considers monthly malaria incidence in three provinces of Mozambique: Niassa, Cabo Delgado, and Nampula,  aggregated at the level of 56 administrative districts from 2017 to 2024. Although arising from a specific epidemiological context, this dataset reflects challenges common to many spatio-temporal applications characterized by areal units observed over relatively long time horizons. The data consist of monthly counts of laboratory-confirmed malaria diagnoses reported by all health facilities in the three provinces, provided by the National Malaria Control Program of Mozambique under a Data Transfer Agreement. Counts were transformed into log-scale monthly incidence per 1,000 inhabitants using district-level population estimates from the 2017 Population and Housing Census and its official projections \cite{ine2017census}. The resulting dataset constitutes a moderately large spatio-temporal panel with 56 spatial units and 89 time points. It exhibits substantial heterogeneity across both space and time, with no missing values. \\

\begin{figure}
    \centering
    \begin{minipage}{0.5\linewidth}
        \centering
        \includegraphics[width=\linewidth]{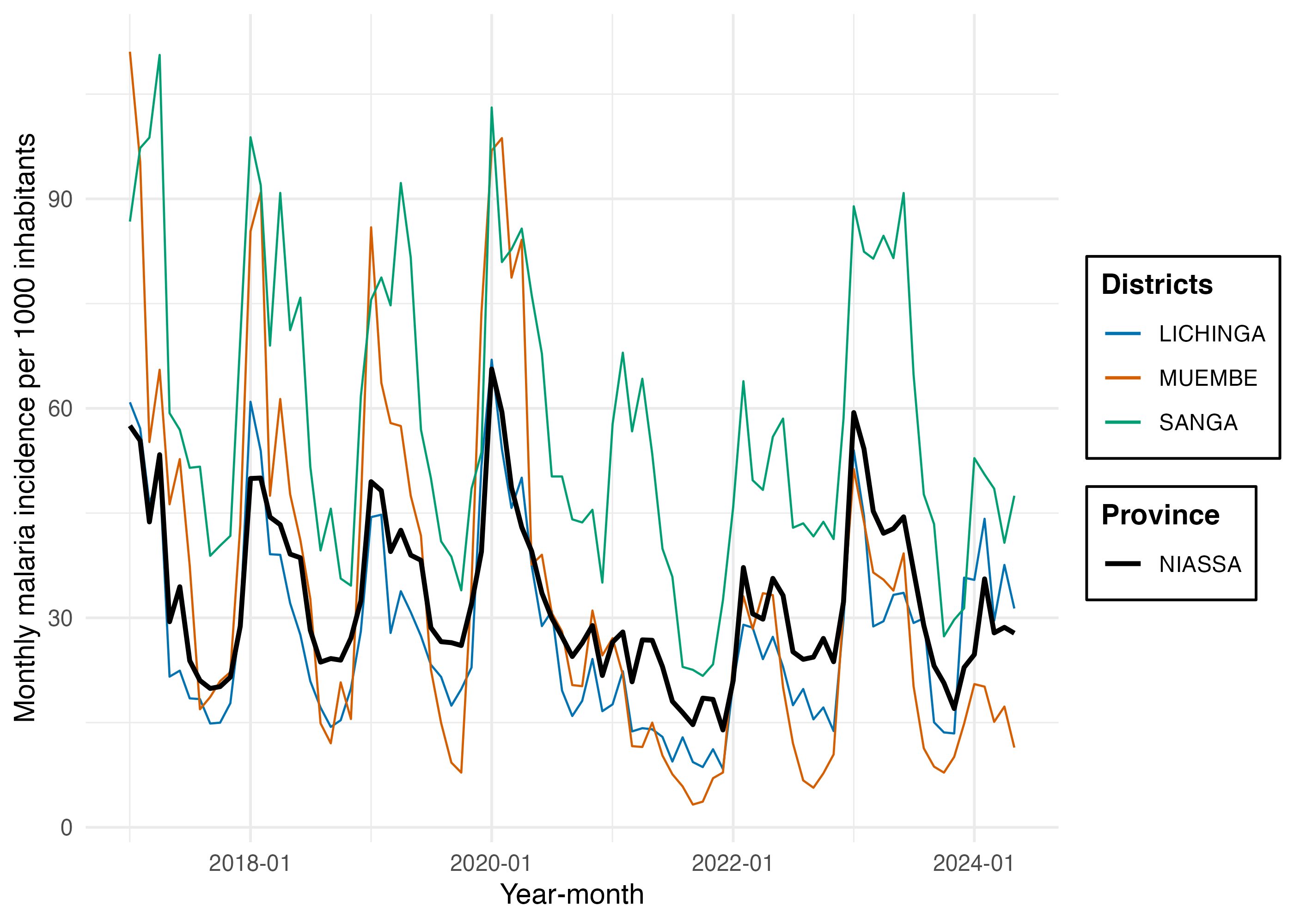}
    \end{minipage}
    \vspace{0.5cm}
    \begin{minipage}{0.5\linewidth}
        \centering
        \includegraphics[width=\linewidth]{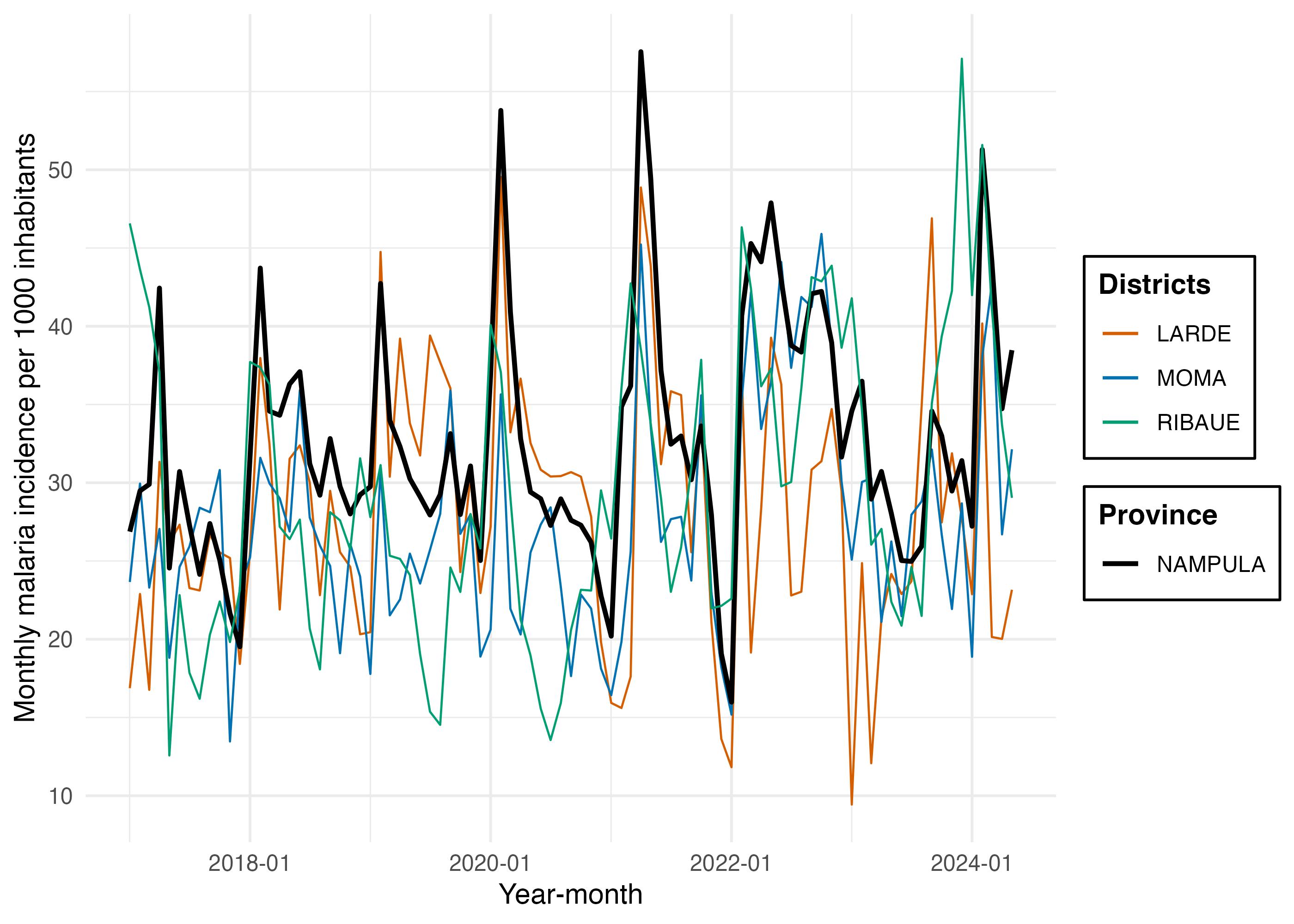}
    \end{minipage}
    \vspace{0.5cm}
    \begin{minipage}{0.5\linewidth}
        \centering
        \includegraphics[width=\linewidth]{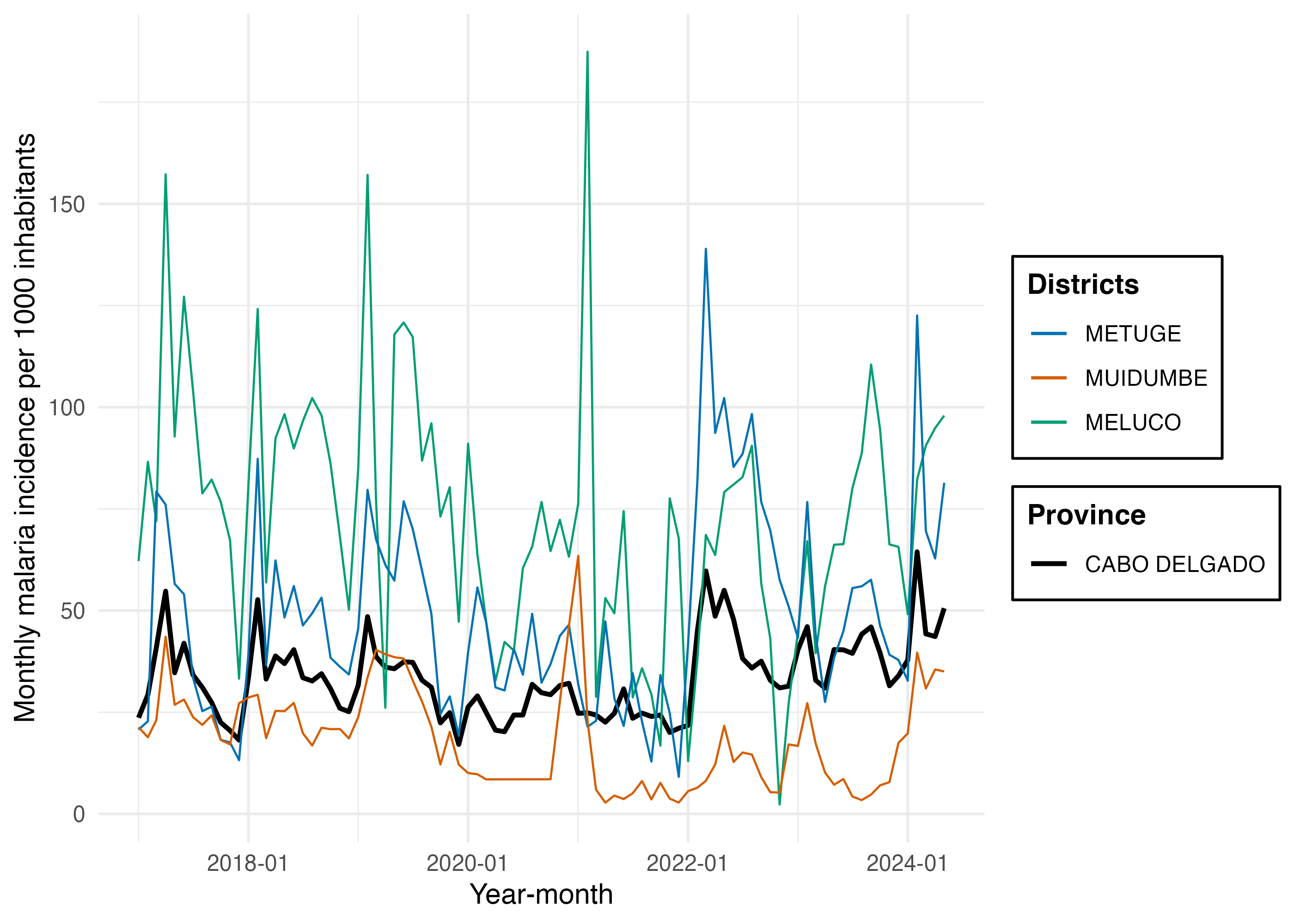}
    \end{minipage}
    \caption{Monthly observed malaria incidence per 1,000 inhabitants from January 2017 to June 2024 for selected districts in the provinces of Niassa, Nampula, and Cabo Delgado. Each panel displays province-level incidence together with representative district trajectories.}
    \label{fig:temporal_trend_combined}
\end{figure}

\begin{figure}
    \centering
    \includegraphics[width=0.5\linewidth]{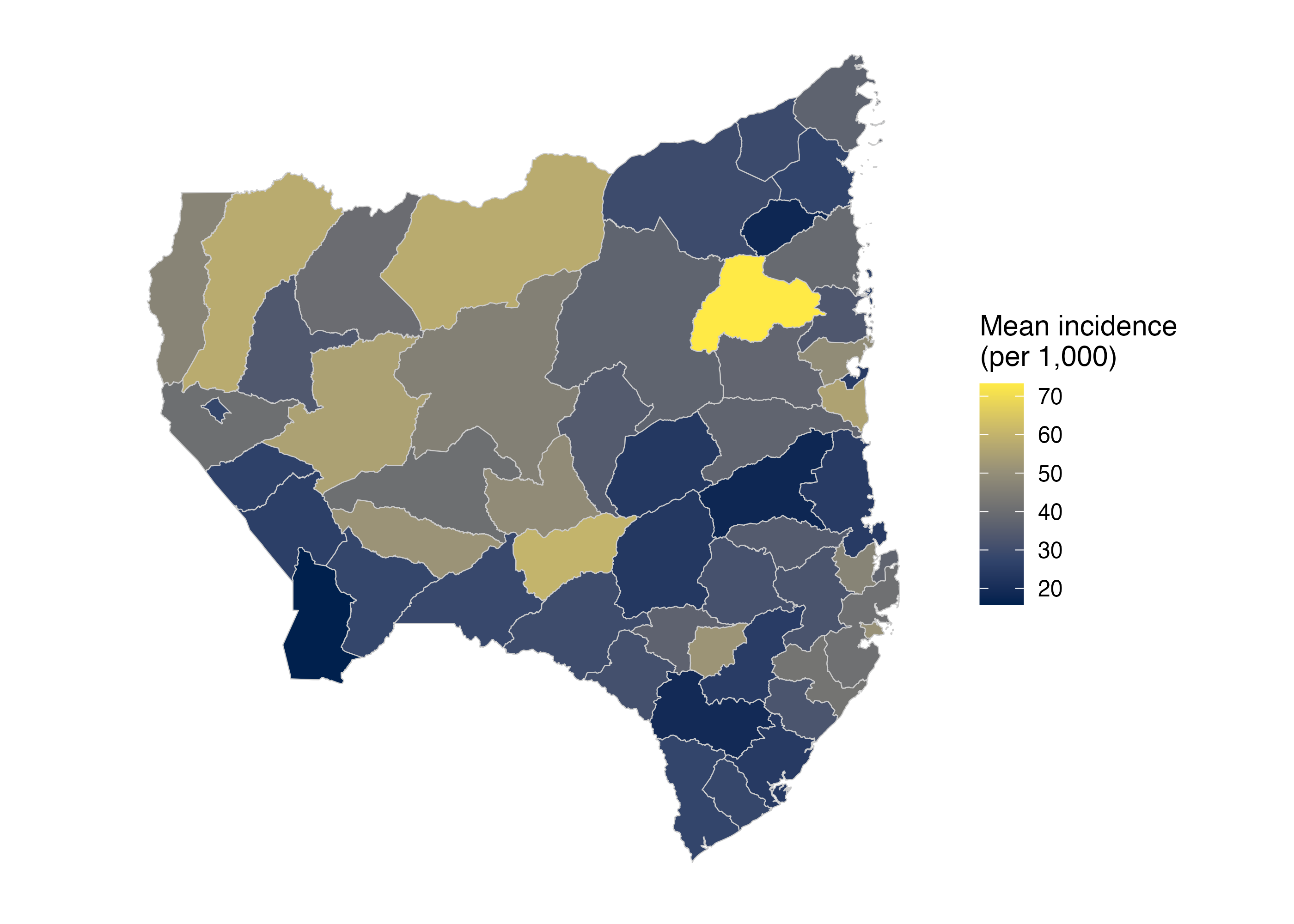}
    \caption{Average observed malaria incidence across all the districts of the provinces of Niassa, Nampula and Cabo Delgado between 2017 and 2024.}
    \label{fig:spatial_distrib}
\end{figure}

\begin{figure}[h]
\centerline{\includegraphics[width=0.5\linewidth]{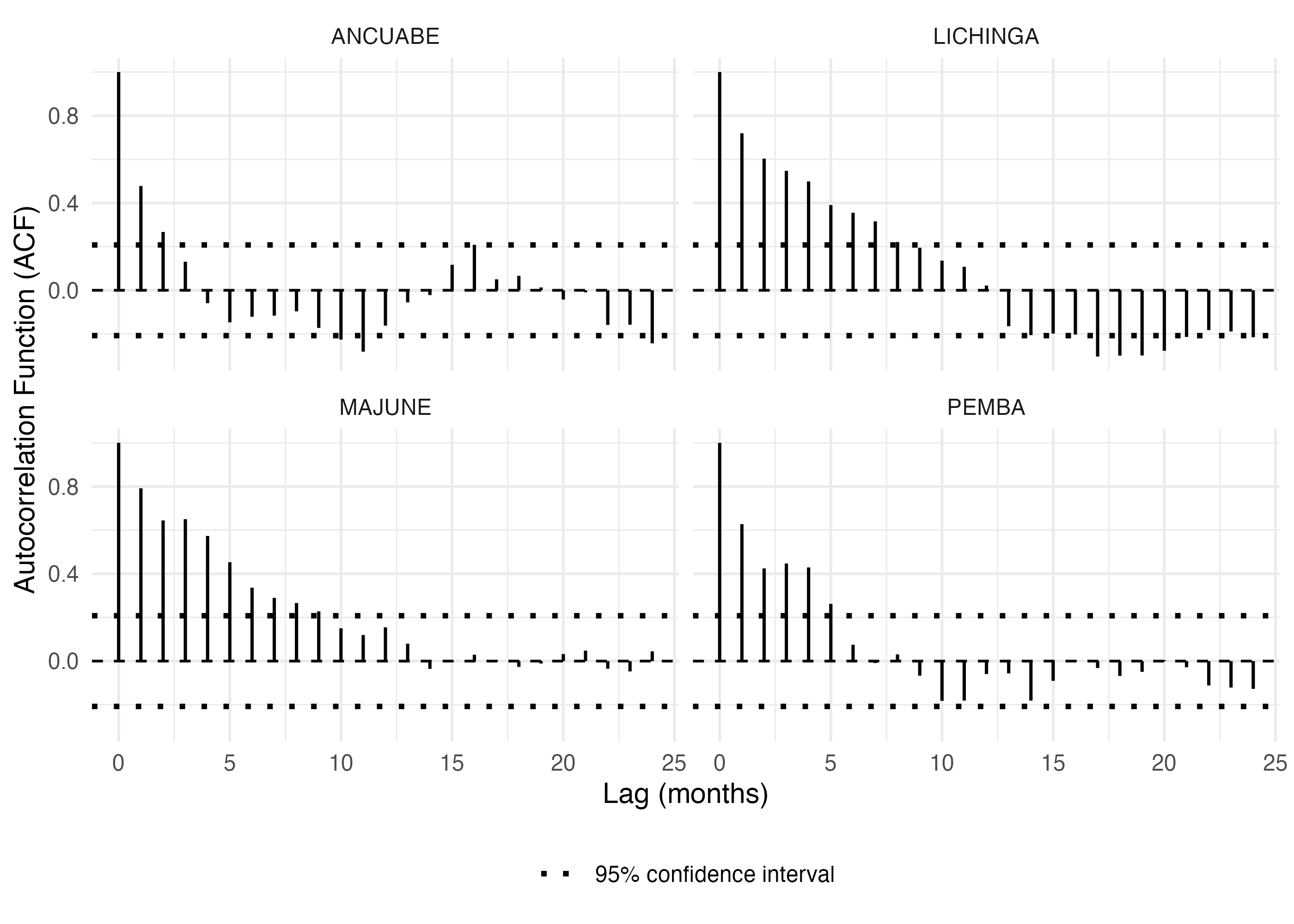}}
\caption{Autocorrelation functions (ACFs) of monthly malaria incidence for the districts of Ancuabe, Luchinga, Majune and Pemba with approximate 95\% confidence (dotted line).\label{fig1.appA.ex1}}
\end{figure}

\begin{figure}[h]
\centering
\begin{minipage}{0.48\textwidth}
    \centering
    \includegraphics[width=\linewidth]{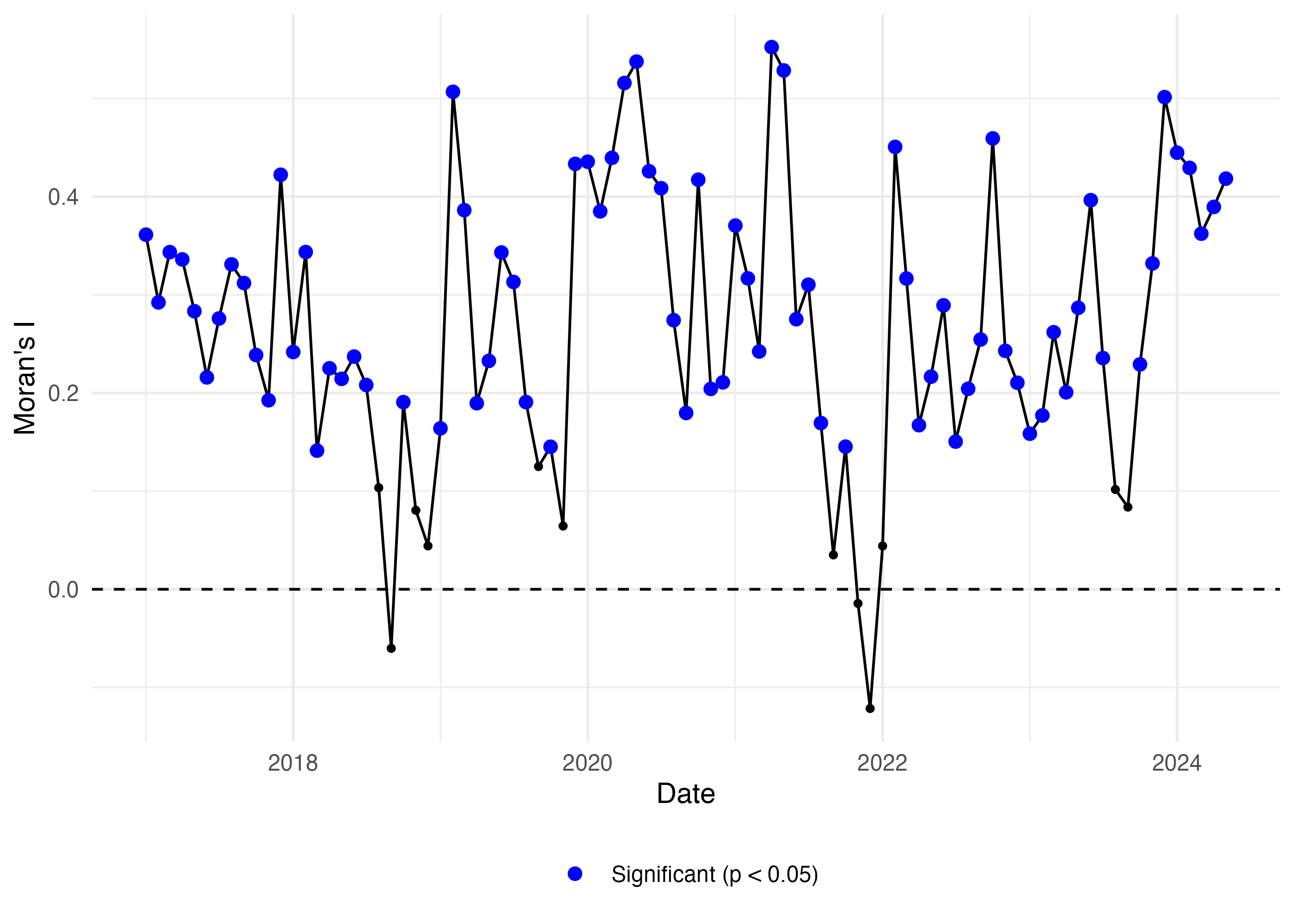}
    (a)
\end{minipage}
\hfill
\begin{minipage}{0.48\textwidth}
    \centering
    \includegraphics[width=\linewidth]{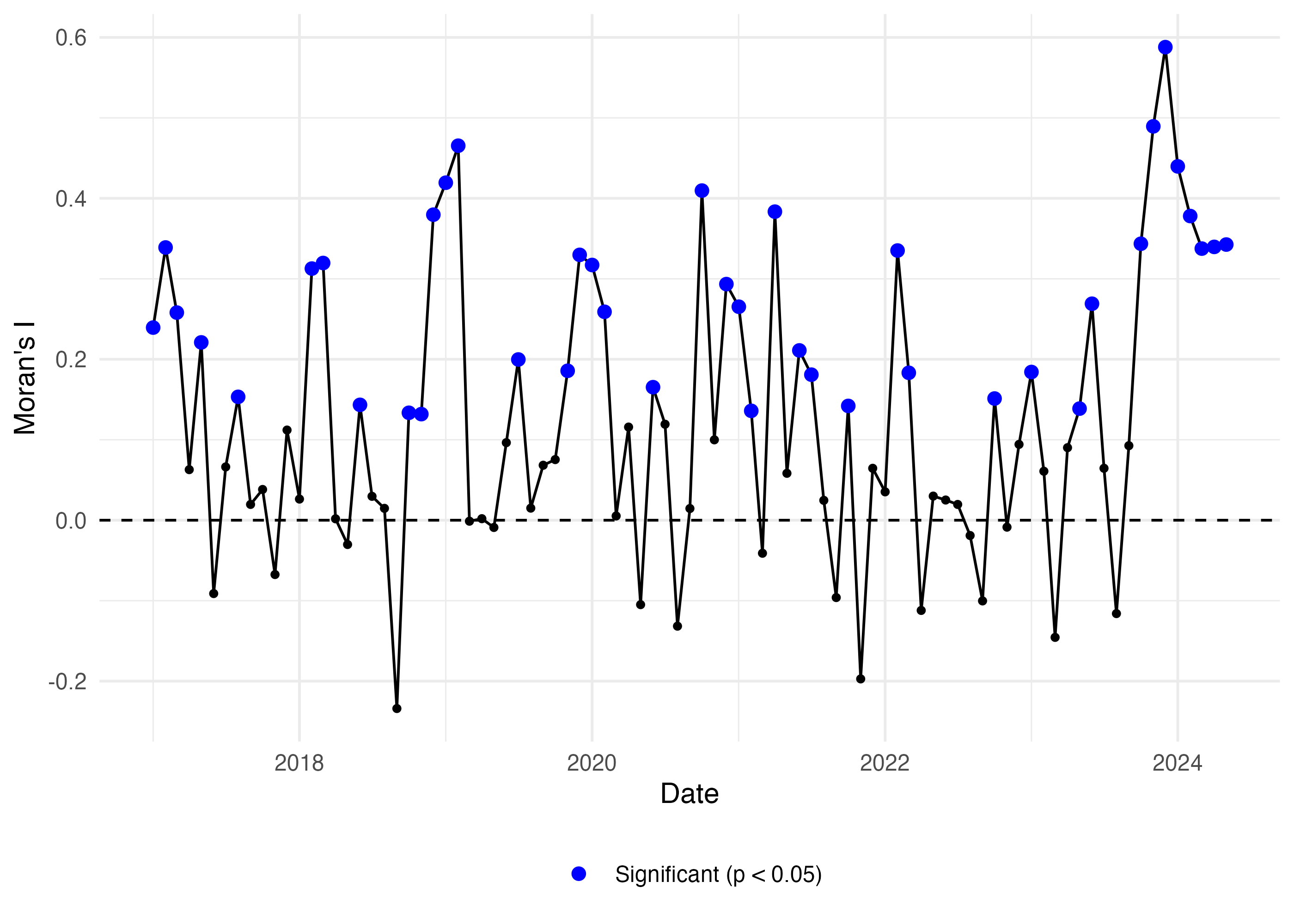}
    (b)
\end{minipage}
\caption{Temporal evolution of Global Moran's I computed from model residuals. 
(a) Pearson residuals from a model including district-specific intercepts and seasonal fixed effects. (b) Residuals from a model including district-specific intercepts, seasonal fixed effects, and independent district-specific Gaussian processes. Blue points indicate months for which Moran's I was statistically significant based on permutation tests ($p<0.05$).}
\label{fig:moran_residuals}
\end{figure}

Figure \ref{fig:temporal_trend_combined} illustrates the temporal evolution of monthly incidence in selected districts within each province. All districts exhibit marked seasonality, consistent with malaria transmission dynamics, yet their temporal trajectories differ in baseline risk, amplitude, and persistence of peaks. Aggregation at the province level smooths over this heterogeneity, potentially obscuring high-risk districts. Furthermore, Figure \ref{fig:spatial_distrib} illustrates the spatial distribution of the average incidence over the full study period, revealing geographic clustering, with neighboring districts tending to exhibit similar levels of risk. Exploratory analyses further confirm the presence of residual temporal and spatial dependence after accounting for the main systematic components. To assess temporal dependence, we examined autocorrelation functions plots computed from Pearson residuals of a model including district-specific intercepts and seasonal effects (Figure \ref{fig1.appA.ex1}). We observed positive correlation at short lags and persistent dependence beyond immediate months, a consistent pattern across districts. To assess residual spatial dependence, we examined the temporal evolution of Global Moran’s I computed from the same Pearson residuals used in the autocorrelation analysis (Figure \ref{fig:moran_residuals}, panel a). The figure shows that Moran's I values are predominantly positive throughout the study period and were statistically significant for approximately 75\% of the months, indicating that substantial spatial correlation persists even after accounting for fixed effects. \\

To further investigate the remaining spatial dependence, we fitted independent district-specific time-series models that included an intercept, seasonal effects and random effects from a temporal Gaussian process. We recomputed Global Moran's I using the resulting residuals. Incorporating independent district-specific temporal Gaussian processes reduced the magnitude of the residual spatial dependence, with the residuals' Moran's I remaining significant for approximately 50\% of the time points (Figure \ref{fig:moran_residuals} panel b). Although temporal random effects explain an important proportion of the residual variability, considerable structure remains, suggesting that independent temporal processes alone are insufficient to fully capture the joint spatio-temporal dependence present in the data.\\

\begin{figure}[ht]
\centering
\begin{minipage}{0.48\textwidth}
    \centering
    \includegraphics[width=\linewidth]{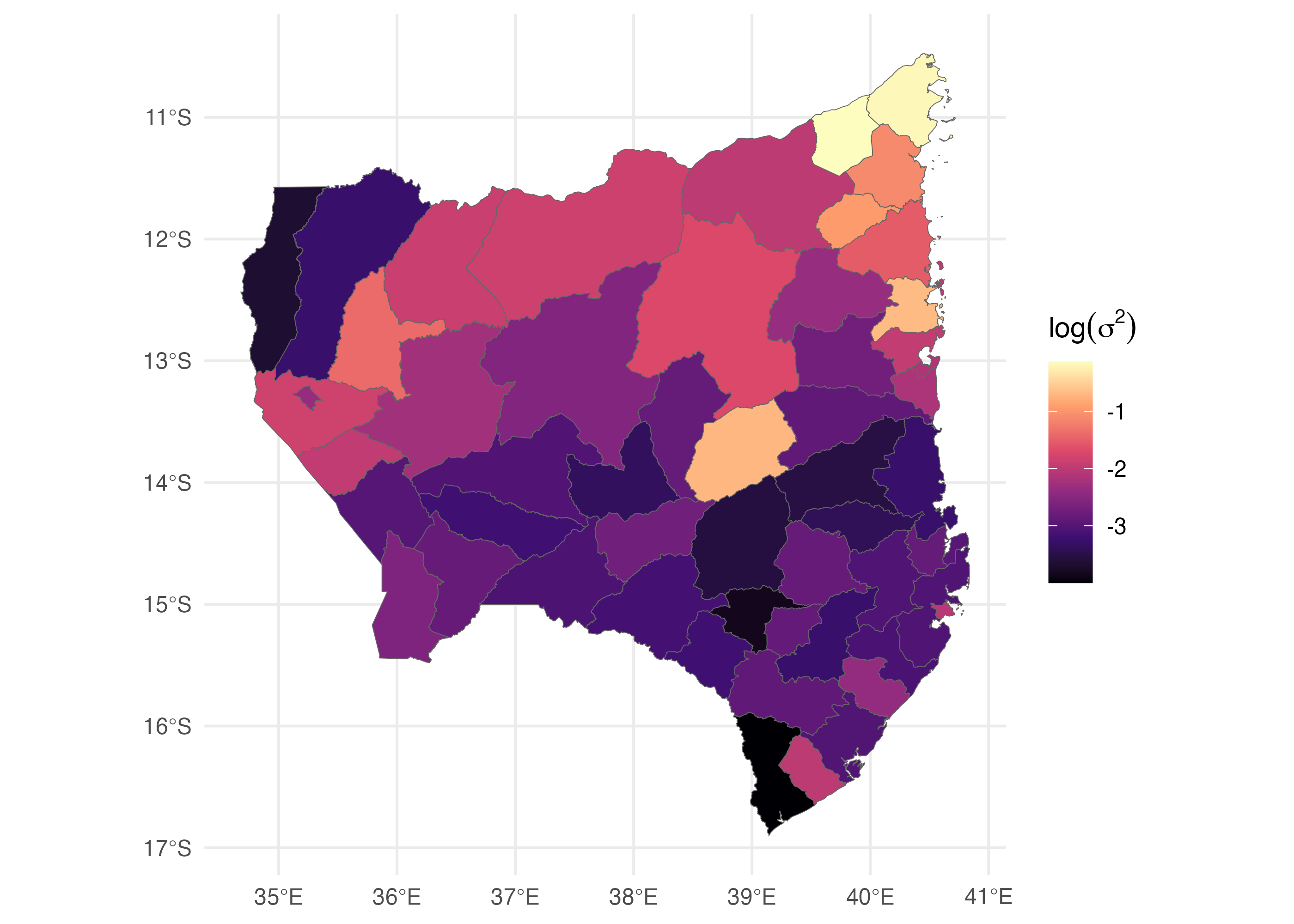}
    (a) Estimated district-specific $\log(\sigma^2)$.
    \label{fig:logsigma_map}
\end{minipage}
\hfill
\begin{minipage}{0.48\textwidth}
    \centering
    \includegraphics[width=\linewidth]{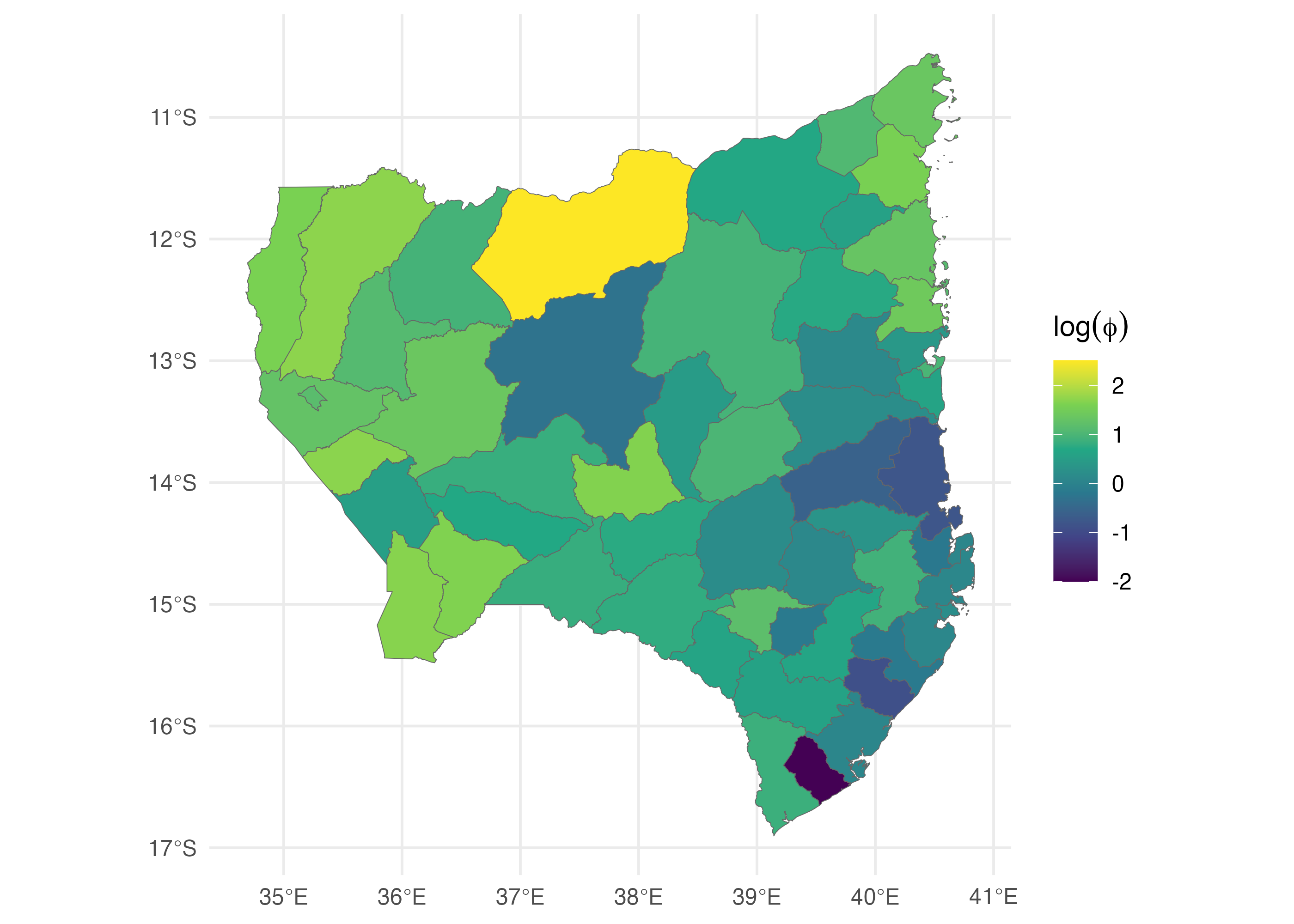}
    (b) Estimated district-specific $\log(\phi)$.
    \label{fig:logphi_map}
\end{minipage}
\caption{Estimated temporal Gaussian process hyperparameters obtained by fitting independent Gaussian process models to each district.}
\label{fig:gp_hyperparameters}
\end{figure}

Additionally, we investigated whether characteristics of the temporal dynamics exhibit any kind of spatial structure. To this end, we examined district-specific estimates of the variance and range hyperparameters obtained from the independently fitted Gaussian processes models. Figure \ref{fig:gp_hyperparameters} displays the estimated values of the temporal hyperparameters, $\log(\widehat{\sigma^2})$ and $\log(\widehat{\phi})$ across districts, revealing clear spatial patterns in both parameters. To quantify the extent of the spatial dependence of the hyperparameters of the GP, BYM2 models \cite{bym2} were fitted separately to the estimated values of $\log(\sigma^2)$ and $\log(\phi)$. The posterior mean of the BYM2 mixing parameter was 0.58 (0.38,0.78) and 0.41 (0.15,0.56) for  $\log(\widehat{\sigma^2})$ and $\log(\widehat{\phi})$ respectively, indicating that a substantial proportion of the latent variability in both temporal hyperparameters is spatially structured. These findings suggest that neighboring districts tend to exhibit similar temporal variability and persistence, providing empirical support for introducing spatially structured priors on the GP hyperparameters. \\ 

Collectively, these exploratory analyses indicate that malaria incidence exhibits heterogeneous temporal trajectories across districts, persistent residual temporal dependence, residual spatial autocorrelation, and spatially structured temporal dynamics. A key inferential challenge is therefore to develop models capable of accommodating district-specific temporal evolution while borrowing strength across neighboring regions through spatially correlated temporal characteristics. These observations directly motivate the modeling framework proposed in the following sections.

\section{Existing spatio-temporal models for areal data}
\label{Section.models}
We provide a summary of spatio-temporal models for areal data that draws primarily from existing reviews \cite{review1_st, review2_st, lopez2009review, macnab2022bayesian_revision}. It is not our intention to provide an in-depth technical description or overview of these models.  Readers interested in detailed formulations are encouraged to consult these references directly. \\

We begin with common notation. Let a study area be divided into \textit{I} non-overlapping regions, for example districts, counties, or municipalities, indexed by $i = 1, ..., I$. The temporal dimension is indexed by $t = 1, ..., T$, representing time points such as days, weeks, or years. We denote the response variable at location $i$ and time $t$ by $Y_{it}$, with observed value $y_{it}$. The data are typically assumed to be conditionally independent given latent effects and covariates and to follow a probability distribution, often within the exponential family. The mean $\mu_{it}$ may be linked to a linear predictor through a link function $g(\cdot)$, so that $\eta_{it}=g(\mu_{it})$. The linear predictor is generally expressed additively as a sum of components capturing distinct sources of variability. \\

A broad class of spatio-temporal models can be represented by the following general form for the linear predictor, $\eta_{it}$:
\begin{equation}
\eta_{it} = \beta_0 + \mathbf{X}_{it} \boldsymbol{\beta} + S_{i} + D_{t} + Z_{it} + \epsilon_{it}.
\label{eq.generalmodel}
\end{equation}
Here, $\beta_0$ represents the intercept, $\mathbf{X}_{it}$ is a design matrix containing covariates, and $\boldsymbol{\beta}$ denotes a vector of regression coefficients. The terms $S_i$, $D_t$, and $Z_{it}$ represent spatial, temporal, and spatio-temporal components, respectively. The term $\epsilon_{it}$ corresponds to the unstructured heterogeneity term. For Gaussian linear models, $\epsilon_{it}$ represents a convolution of unstructured extra-variability and Gaussian noise; for Poisson or binomial models, where the distributional error is a direct function of the mean, and therefore not separately estimated, $\epsilon_{it}$ only captures unstructured extra-variability. \\

Depending on how the spatio-temporal effects are specified, models can broadly be classified into three groups: (i) additive spatial and temporal effects without interaction, (ii) models with both additive and explicit interaction terms, and (iii) models that include only a spatio-temporal interaction component. Because these models are typically hierarchical with intractable likelihoods, inference is usually conducted under the Bayesian paradigm, and much of the literature focuses on prior specifications of the latent effects.\\

In the first group, spatial and temporal effects enter additively, so that $Z_{it}$ is omitted from equation~(\ref{eq.generalmodel}). For the spatial component $S_i$, common choices include conditional autoregressive (CAR) priors, often combined with an unstructured Gaussian term to form the Besag-York-Mollié (BYM) model \cite{besag1991bayesian}, the Leroux prior \cite{leroux2000estimation}, adaptive CAR structures \cite{macnab2023adaptive}, and spatial autoregressive (SAR) priors \cite{anselin1992spatial}. Temporal components $D_t$ are typically modeled using autoregressive (AR) processes, first- or second-order random walks (RW), or, less commonly, temporally structured CAR priors. For example, Napier et al. \cite{napier2016model} specify a CAR model for spatial effects at each time point together with a common first-order random walk for the temporal trend.\\

A second class of models explicitly incorporates a spatio-temporal interaction $Z_{it}$. A widely cited framework is that of Knorr-Held \cite{knorr2000bayesian}, which decomposes both $S_i$ and $D_t$ into structured (CAR and RW) and unstructured Gaussian components. The interaction term $Z_{it}$ can be specified in four different forms, depending on whether structured or unstructured spatial effects interact with structured or unstructured temporal effects. Sahu and Böhning \cite{sahu2022bayesian} propose a more parsimonious version that retains structured effects for $S_i$ and $D_t$ while modeling $Z_{it}$ using AR, RW, or CAR processes. Other approaches combine CAR models for $S_i$ with spline-based temporal trends for $D_t$ and spatially varying splines for $Z_{it}$ \cite{macnab2007spline, macnab2011gaussian}. Alternatively, tensor-product spline representations have been used to model smooth spatio-temporal surfaces \cite{ugarte2017one, feng2022spatial}. \\

A third group of models includes only the spatio-temporal interaction $Z_{it}$ in the linear predictor, omitting separate $S_i$ and $D_t$ terms. Rushworth, Lee and Mitchell \cite{rushworth2014spatio} model $Z_{it}$ using multivariate autoregressive processes with spatially structured precision matrices. Rushworth et al. \cite{rushworth2017adaptive} and Gao and Bradley \cite{gao2019bayesian} propose adaptive CAR structures where the spatial adjacency matrix is estimated from the data. In contrast, Macnab \cite{macnab2018some}, Corpas et al. \cite{corpas2020use} and Reich and Hodges \cite{reich2008modeling} use fixed adjacency matrices but introduce spatially varying scale or precision parameters to modulate local dependence. Macnab \cite{macnab2023adaptive} provides an extensive discussion of these models and their extensions. \\

While existing spatio-temporal models provide principled mechanisms for borrowing information across space and time, they also share structural features that motivate the development of alternative formulations. In the first group of models, where spatial and temporal components enter additively, the temporal evolution is driven by a single process that is common to all regions  capturing the overall evolution of risk, and spatial effects primarily shift the baseline level across areas. In such settings, regions may differ in their average risk but tend to evolve in parallel over time, which may limit the ability to capture localized temporal dynamics. Models in the second group introduce explicit spatio-temporal interaction terms that allow regions to deviate from the global temporal trend. Nevertheless, the temporal structure in these models typically remains anchored by a common temporal component, with the interaction term capturing deviations around this global process rather than defining fully distinct temporal trajectories for each region. The third group of models offers greater flexibility as spatial dependence is generally imposed directly on the latent spatio-temporal effects, and the parameters governing temporal dependence such as autoregressive coefficients or smoothing parameters are typically shared across regions. Consequently, information sharing across space occurs primarily through the latent risk surface rather than through the parameters governing temporal dynamics themselves. As a result, neighboring regions may exhibit similar risk levels while still being constrained to share common assumptions about temporal dependence. \\

An important step toward relaxing the reliance on globally shared temporal dynamics and common temporal dependence parameters is the adaptive spatio-temporal Gaussian Markov Random Fields (GMRF) model of Macnab \cite{macnab2023adaptive}, which allows aspects of temporal smoothing to vary over space, shifting attention from spatially structured outcomes to spatially structured temporal behavior. Building on this idea, the next section introduces a model that places spatial dependence directly on the parameters governing local temporal dynamics, allowing each region to follow a flexible time process while still borrowing strength from nearby areas. This yields a framework that lies between fully shared temporal trends and completely independent regional time series, thereby broadening the class of available spatio-temporal models for areal data. \\

\section{A time-series model for areal data using area-specific Gaussian processes with spatially correlated hyperparameters}
\label{Section.male}
Classical spatio-temporal models exploit correlation by borrowing strength across neighboring areas and time points, which can improve stability of estimation in settings with sparse or noisy observations. In this work, we propose a Bayesian hierarchical modeling framework that retains the benefits of borrowing information across space and time while allowing greater adaptability in how temporal changes are modeled within each location. In particular, temporal trajectories are modeled in a way that allows area-specific evolution while maintaining spatial structure in the parameters governing temporal variability. This formulation provides a flexible representation of spatio-temporal processes while retaining the hierarchical structure commonly used in disease mapping and related applications. The approach we propose follows the spirit of Macnab \cite{macnab2023adaptive}, which introduces spatial adaptivity through structured precision parameters, but we extend this idea to Gaussian-process time series. \\

Building upon the notation introduced in Section \ref{Section.models}, we now present a data-driven spatio-temporal modeling framework, with the corresponding linear predictor given by, 
\begin{equation}
\eta_{it} = \mathbf{X}_{it}\boldsymbol{\beta}_i + Z_{it} + \epsilon_{it},
\label{eq.male}
\end{equation}
where $\mathbf{X}_{it}$ denotes the design matrix containing the covariates of interest, $\boldsymbol{\beta}_i$ is a vector of area-specific regression coefficients, and $Z_{it}$ denotes the realization at time $t$ of a area-specific temporal Gaussian process,  $\boldsymbol{Z}_i = (Z_{i1},\ldots,Z_{iT})$. Where $\boldsymbol{Z}_i$ is assumed to be stationary with covariance, 
\begin{equation}
     \gamma_i(u) = cov(Z_{it}, Z_{it'}) = \sigma_i^2 C_i(u,\phi_i),
     \label{Eq.covS}
\end{equation}
where $u=|t-t'|$ denotes the temporal lag, $\sigma_i^2$ is the marginal variance, and $C_i(\cdot, \phi_i)$ is a valid correlation function governed by the temporal range parameter $\phi_i$. We adopt Gaussian processes for temporal modeling because they provide a flexible, continuous-time representation of temporal dependence. The temporal structure is encoded directly through the covariance function, with commonly used kernels such as the Matérn, exponential, and squared exponential families, allowing the model to represent smooth, persistent, or rapidly decaying dependence patterns as supported by the data.\\

The exploratory analyses presented in Section \ref{Section.examples} indicated that neighboring regions not only exhibit similar disease risk but also share similar temporal characteristics, with the estimated Gaussian-process hyperparameters displaying clear spatial dependence. Motivated by these findings, we place spatial structure on the hyperparameters of the Gaussian processes rather than directly on the outcomes. Specifically, we assume that: (i) the log-transformed temporal variance parameters $(\log(\sigma^2_1),$ $ \ldots,$ $\log(\sigma^2_I))$ follow a CAR prior using a Queen contiguity matrix to define spatial adjacency; and (ii) the log-range parameter $\log(\phi_i)$ follows a Gaussian distribution conditionally on $\log(\sigma^2_i)$,  
$$\log(\phi_i)| \log(\sigma^2_i)\sim \mathcal{N}(\mu_{\log(\phi)} + \gamma_{\log(\phi),\log(\sigma^2)}(\log(\sigma^2_i)), \chi^2),$$ 
where $\mu_{\log(\phi)}$ controls the overall level of the temporal range, $\gamma_{\log(\phi),\log(\sigma^2)}$ captures the association between temporal variability and temporal range, and variance $\chi^2$ represents the residual between-district variability in the range-parameter not explained by the GP-variances parameters. The prior for $\mu_{\log(\phi)}$ $\left( \mu_{\log(\phi)} \sim N(\log(0.35),0.4^2)\right)$ centered the temporal range around a plausible value while allowing moderate uncertainty; the prior for $\gamma_{\log(\phi),\log(\sigma^2)}$ $\left( N(0,0.5^2)\right)$ was centered at zero so that no association between temporal variance and temporal range was assumed a priori; and the prior for $\chi^2$ controls the residual between-district variability in the log-range parameters after accounting for their dependence on the GP variance parameters.\\

This hierarchical specification reflects the idea that regions may differ in how their outcomes evolve over time, while geographically proximate areas are likely to exhibit related temporal dynamics due to shared environmental, demographic, or epidemiological factors. Rather than imposing spatial dependence directly on the latent outcomes, the model introduces spatial structure on the parameters governing temporal dependence. In this way, neighboring regions can share information about the characteristics of their temporal dynamics, while still allowing for area-specific temporal trajectories. The approach follows the spirit of Macnab \cite{macnab2023adaptive}, which introduces spatial adaptivity through structured precision parameters, but we extend this idea to Gaussian-process time series. \\

In this developmental study, we adopt a Gaussian working likelihood applied to suitably transformed outcomes in order to focus on the proposed spatio-temporal structure. Specifically, we assume that the transformed response follows a normal distribution with additive Gaussian noise $\epsilon_{it}\sim\mathcal{N}(0,\tau_i^2)$ representing measurement error and, possibly, small-scale variability not captured by the structured components of the model. Note that the parameter $\tau_i^2$ is also known as nugget effect. This distributional specification corresponds to a latent Gaussian representation on the canonical link scale of generalized linear models and provides a computationally tractable framework for developing and validating the proposed hierarchical spatio-temporal dynamics. Extensions to fully discrete likelihoods (e.g., Poisson, binomial or Negative binomial) are conceptually straightforward and follow directly from the same latent structure. \\

We assign independent weakly informative priors to the area-specific's nugget effect by specifying $\log(\tau_i^2) \sim N(log(0.5),1)$. The prior mean was chosen to reflect the magnitude of residual variation observed in the exploratory district-specific analyses, while the relatively large variance provides sufficient flexibility for the data to determine the nugget effect. For the regression coefficients, we use a non-informative multivariate Gaussian prior, $\boldsymbol{\beta}_i \sim \mathcal{N}_{p+1}(0, 10^{5}I_{p+1})$,  where $p$ denotes the number of covariates and $I_{p+1}$ is the identity matrix of dimension $p+1$. Under this specification, each area $i$ is characterized by parameters $\theta_i = (\boldsymbol{\beta}_i, \sigma^2_i, \phi_i, \tau^2_i)$. \\

Note that when no spatial correlation is assumed in the variance and range parameters of the Gaussian processes, the model of equation (\ref{eq.male}) reduces to a model with independent time series for each region, which will be referred to as the \textit{independent GP time-series model}. 
\subsection{Model estimation, prediction and implementation}
\label{Sect.male.est}
Posterior inference was performed within a Bayesian framework using Markov chain Monte Carlo (MCMC) sampling because the joint posterior distribution does not admit a closed-form expression. The parameter estimation relies on iterative sampling from the posterior distribution. During model development, we derived and implemented a tailor-made MCMC algorithm based on the full conditional distributions, employing different sampling strategies (Metropolis-adjusted Langevin algorithm (MALA) \cite{mala, roberts1996exponential}, Metropolis-Hastings \cite{metropolis1953, hastings1970}, and Gibbs Sampling \cite{gelfand1990}) according to the analytical form of each conditional posterior. This implementation was used to verify the estimation procedure and is described in detail in Appendix A. For the analyses presented in this paper, however, the model was fitted using the software for Bayesian analysis \texttt{Stan} \cite{stan2015} (version 2.38) through the \texttt{cmdstanR} package (version 0.9.0) \cite{cmdstan}, which provides substantially more efficient computation through Hamiltonian Monte Carlo, more precisely using the No-U-Turn Sampler (NUTS) \cite{NUTS}. \\

Equation \ref{mod.malaria} describes the model structure used to fit the proposed model to the \textit{malaria data}:  
\begin{equation}
\begin{aligned}
\log \left(Y_{it}\right) &= \eta_{it}, \\
\eta_{it} &= \beta_{0i} + \beta_{1i} \sin\left( \frac{2\pi t}{12} \right) + \beta_{2i} \sin\left( \frac{2\pi t}{6} \right) \\
&+ \beta_{3i} \cos\left( \frac{2\pi t}{12} \right) + \beta_{4i} \cos\left( \frac{2\pi t}{6} \right) + Z_{it} + \epsilon_{it},
\end{aligned}
\label{mod.malaria}
\end{equation}
where $Y_{it}$ denotes the monthly malaria incidence in district $i$ at month $t$, and $Z_i$ a Gaussian process with a Matérn correlation function \cite{matern2013spatial}. The Matérn family is widely adopted in spatial and spatio-temporal statistics because it provides a flexible and interpretable representation. We fixed the smoothness parameter at $\kappa = 1.5$, a commonly adopted choice, because it provides a balance between flexibility and computational simplicity.  Since this work focuses on the Gaussian distribution for methodological development, we model the log-transformed malaria incidence. This transformation corresponds to the canonical link function of a Poisson generalized linear model, and yields approximately variance-stabilized and symmetric responses. Consequently, the proposed specification can be interpreted as a latent Gaussian representation on the link scale of the underlying non-Gaussian data-generating processes. \\

As explanatory variables, we include first- and second-order Fourier harmonics to capture large-scale seasonal patterns. The number of harmonics was selected based on the Akaike Information Criterion (AIC) values from simpler linear models with varying Fourier terms. This choice is motivated by the strong and regular annual seasonality observed in malaria incidence, which is well approximated by a small number of sinusoidal components and allows for a parsimonious and interpretable representation of temporal effects. \\

We run the MCMC algorithm for 2,000 iterations, which includes a burn-in period of 1,000.  Trace plots, autocorrelation plots, and Geweke running plots were used for visual inspection of the convergence, in addition to the Geweke diagnostic as a formal assessment. Furthermore, the data were partitioned into training and test sets. The training set covered the period from January 2017 to September 2023, and the test set comprised data from October 2023 to June 2024. This division was designed to evaluate the out-of-sample predictive performance of the model.

Posterior predictive inference is based on the predictive distribution
$$
p(\boldsymbol{y}^{\text{new}} \mid \boldsymbol{y}) = \int p(\boldsymbol{y}^{\text{new}} \mid \boldsymbol{\theta}) p(\boldsymbol{\theta} \mid \boldsymbol{y}) \, d\boldsymbol{\theta},
$$
Predictive samples are generated by drawing new observations conditional on posterior samples of the model parameters, thereby accounting for both parameter and observational variability.\\

\subsection{Model performance evaluation}
\label{Sect.male.performance}
To assess the performance of our proposed model, we compared it against four alternatives, each representing a set of important variations in modeling approaches found in the literature: (M1) the spatio-temporal model with space-time interaction by Knorr-Held \cite{knorr2000bayesian}, we selected the model with the interaction term that produced the lowest WAIC; (M2) the spatio-temporal model of Rushworth et al. of the first order \cite{rushworth2014spatio}; (M3) the spatio-temporal model of Rushworth of the second order \cite{rushworth2014spatio}; and (M4) the independent GP time-series model. These benchmark models were selected based on their theoretical relevance and the availability of public software implementations, enabling their use across our case study as well as in applied research more generally. \\

Models in (M1) were implemented using the \texttt{R-INLA} package \cite{INLAsoft}, model (M2) and (M3) were implemented using the \texttt{CARBayesST} package \cite{lee2016}. The independent GP time-series model was implemented using the \texttt{PrevMap} package \cite{giorgi2017prevmap}. Note that although \texttt{PrevMap} has originally been developed for geostatistical modeling using spatial coordinates, it can be adapted for time series applications by treating time as a one-dimensional coordinate and pairing it with a constant vector of ones instead of spatial coordinates. Other models available in \texttt{CARBayesST}, such as those proposed by Napier et al. \cite{napier2016model} and Lee and Lawson \cite{lee2016}, were not considered because they are formulated for discrete outcome distributions (e.g., Poisson or binomial), whereas this study deliberately focuses on Gaussian likelihoods for methodological development. Similarly, more recent approaches involving adaptive spatial structures (e.g., \cite{macnab2023adaptive, corpas2020use}) were not considered due to the lack of commonly available implementations in R. For models (M2) and (M3), we ran the MCMC algorithm for 50,000 iterations, with a burn-in period of 10,000 and a thinning interval of 10, yielding 4,000 posterior samples for each parameter. As with our proposed model, we assessed convergence using both graphical and formal diagnostics.  The following subsection provides the detailed specifications of the four models used to benchmark our approach.

\subsubsection{Model specification of benchmark models} 
\label{Sect.benchmark} 
All benchmark models were fitted to the same transformed outcomes as our proposed model to ensure comparability. Model (M1), which applies the approach of Knorr-Held \cite{knorr2000bayesian}, was parametrized as,
\begin{equation} 
\eta_{it} = \beta_0 + \beta_{1i} \sin\left( \frac{2\pi t}{12} \right) + \beta_{2i} \sin\left( \frac{2\pi t}{6} \right) 
+ \beta_{3i} \cos\left( \frac{2\pi t}{12} \right) + \beta_{4i} \cos\left( \frac{2\pi t}{6} \right) + S_i + D_t + Z_{it},
\label{mod.mod.1}
\end{equation}
where $\eta_{it}$ represents the linear predictor for region $i$ and time $t$, and $\beta_{1i},\beta_{2i},\beta_{3i}$ and $\beta_{4i}$ are coefficients associated with the sine and cosine terms. $S_i$ and $D_t$ denote spatial and temporal random effects, respectively, each comprising structured and unstructured components. The spatial effect follows a BYM specification, while the temporal effect consists of a first-order random walk, and an independent Gaussian noise component. In addition, $Z_{it} \sim \mathcal{N}(0, \omega_1^2)$ denotes the space-time interaction term. \\

Models (M2) and (M3), proposed by Rushworth et al. \cite{rushworth2014spatio}, were specified as, 
\begin{equation} 
\eta_{it} = \beta_0 + \sum_{j=1}^2 \left( \beta_{ij}^{(s)} \sin\left( \frac{2\pi j t}{12} \right) + \beta_{ij}^{(c)} \cos\left( \frac{2\pi j t}{12} \right) \right) + Z_{it},
\label{mod.mod.2}
\end{equation}
where $Z_{it}$ represents the spatio-temporal random effect, and the other model components were specified as in Equation \ref{mod.mod.1}. Let $\mathbf{z}_t = (z_{1t}, \dots, z_{It})$ denote the vector of spatial random effects at time $t$. Conditional on the previous time points, $\mathbf{z}_t$ evolves according to a multivariate autoregressive process of either the first (model 2) or second order (model 3). For model (M2), the evolution was given by $\mathbf{z}_t = \rho \mathbf{z}_{t-1} + \epsilon_t$, where $\rho$ is the temporal autoregressive parameter and $\epsilon_t \sim N(0, \omega_2^{-1} \mathbf{Q}(\mathbf{W}, \rho)^{-1})$. Spatial dependence was incorporated via the precision matrix $Q(\mathbf{W}, \rho)$ as defined by Leroux \cite{leroux2000estimation}, where $W$ denotes the neighborhood matrix. The variance of the spatial process is governed by the parameter $\omega_2$. Model (M3) extends this specification by allowing a second-order temporal autoregressive structure.\\

The fourth benchmark model, an independent GP time-series model, assumes spatial independence and models the time series of each region separately. The linear predictor structure followed the structure presented in Equations (\ref{mod.malaria}), but it assumed spatial independence between the area-specific temporal Gaussian processes. This model was implemented using the \texttt{PrevMap} package \cite{giorgi2017prevmap} and estimated via maximum likelihood for computational efficiency. \\

\subsubsection{Comparison metrics}
\label{sect.comp.metrics}
Model performance was assessed using both point-based and distribution-based metrics. For point-based evaluation, we computed the Root Mean Squared Error (RMSE) and Mean Absolute Error (MAE) to quantify discrepancies between observed and predicted values. For distribution-based evaluation, we used the Continuous Ranked Probability Score (CRPS) to evaluate the full posterior predictive distributions, capturing both calibration and sharpness \cite{crps1, crps2}. Additionally, we computed the Empirical Coverage Probability (ECP), defined as the proportion of observed values falling within the model-based credible intervals, to assess the calibration of predictive uncertainty. This was complemented by the median width of the credible intervals, providing a measure of uncertainty dispersion. All metrics were computed for both training and test sets. Notably, unlike RMSE, which rely on point predictions (e.g., mean or median), CRPS evaluates the entire predictive distribution and can be computed at the observation level, enabling both global and local comparisons. Lower CRPS values indicate better predictive performance, reflecting distributions that are both well-calibrated and concentrated around the observed values.

%\backmatter
\section{Results}
\label{Sect.results}

This section presents the results obtained from applying the proposed Gaussian-process time-series model with spatially correlated hyperparameters to monthly malaria incidence data. We first examined convergence diagnostics and posterior computation to assess the stability of the estimation procedure. Model fit and predictive performance were then evaluated by comparing the proposed model with several established spatio-temporal approaches using both point-wise and distribution-based predictive metrics. Finally, we compared the models in terms of the ECP and median prediction interval width to assess the calibration of the model and uncertainty quantification. The main findings are discussed in detail below; however, given the large number of figures and model comparisons, the interpretation and exploration of the results can be further supported through the Shiny application at \url{https://stats73.shinyapps.io/dashboard_stmodel_correhyperpar_vf/}. While the results presented reflect the general behavior observed across the majority of districts, it is important to note that exceptions exist, and model performance may vary for specific districts. \\ 

Both implementations, taylor-made MCMC and Stan's, produced nearly identical posterior summaries and predictive performance, providing a computational validation of the proposed model. This agreement indicates that the inference is robust to the choice of MCMC algorithm. The results presented in this section are based on the Stan implementation. We ran 2,000 MCMC iterations, discarding the first 1,000 as burn-in. Convergence diagnostics based on trace plots and Geweke statistics indicated convergence for most parameters. For some cases, the log-range parameters $\log(\phi_i)$ and the Gaussian-process variance parameters $\log(\sigma_i^2)$ exhibited higher autocorrelation, which is consistent with weak identifiability of Gaussian-process covariance parameters. A detailed discussion of model identifiability is provided in the Appendix B.  Full convergence diagnostics, including trace plots and Geweke statistics for all districts, are available in the accompanying Shiny application. \\

\begin{figure}
    \centering
    \includegraphics[width=0.5\linewidth]{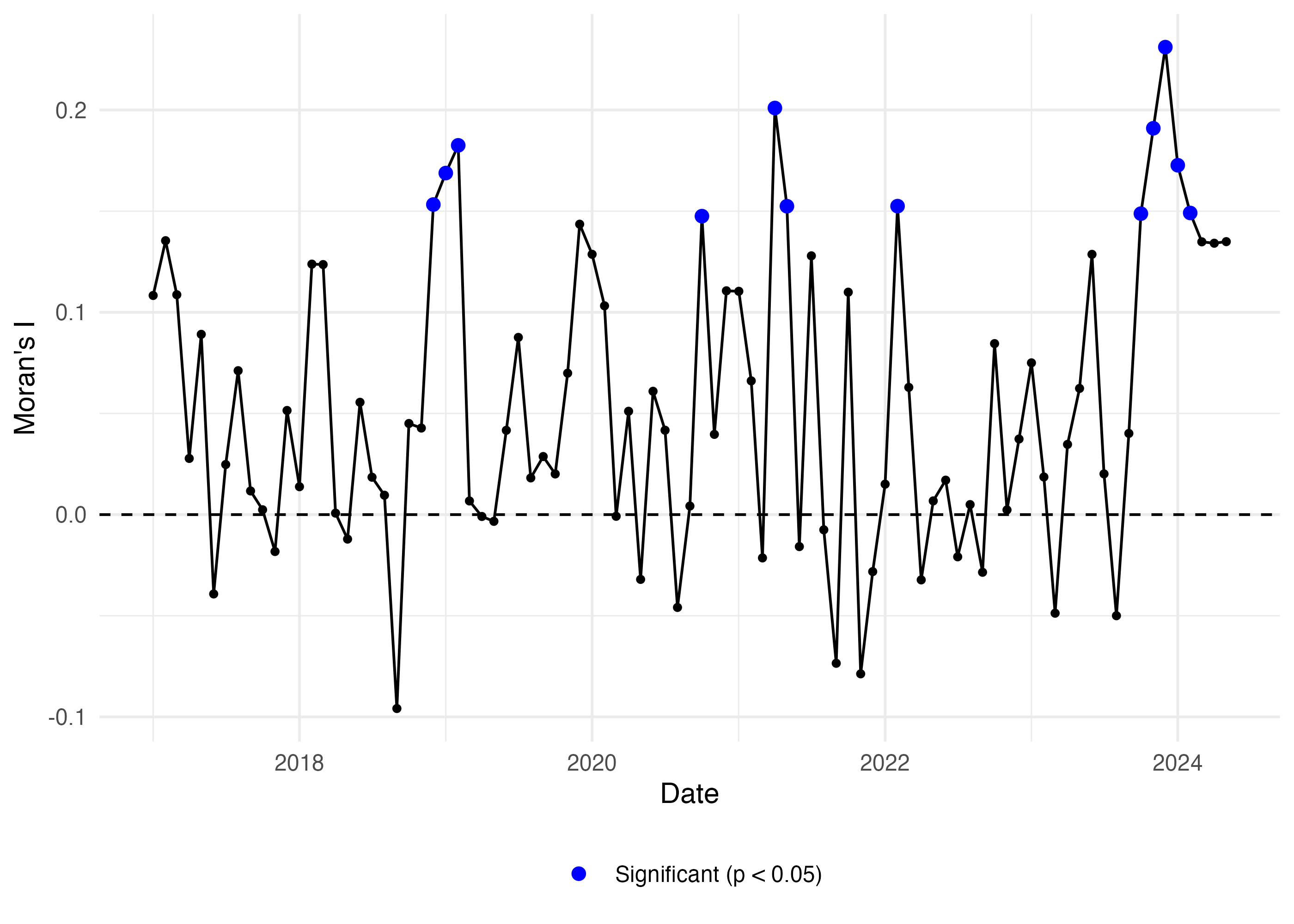}
    \caption{Temporal evolution of Global Moran's I computed from posterior mean residuals of the proposed spatial-temporal model. Blue points indicate months for which Moran's I was significant based on permutation tests (p<0.05).}
    \label{Fig:moran_residuals}
\end{figure}

To assess whether the proposed model adequately captures residual spatial dependence, we computed the global Moran's $I$ statistic on the posterior mean residuals at each time point. Figure~\ref{Fig:moran_residuals} shows that the residual spatial autocorrelation is generally weak throughout the study period, with Moran's $I$ values fluctuating around zero and only a small number of months exhibiting statistically significant positive spatial dependence. Compared with the Moran's $I$ values observed for the raw incidence data (Section~\ref{Section.examples}), both the magnitude and frequency of significant spatial autocorrelation are substantially reduced. This reduction indicates that the spatial structure introduced through the Gaussian-process hyperparameters successfully captures much of the spatial dependence present in the data. The few remaining periods with significant residual spatial autocorrelation suggest the presence of localized spatial patterns not fully explained by the model; however, the overall attenuation of Moran's $I$ provides evidence that the proposed framework effectively accounts for the dominant spatial dependence while retaining flexibility in the temporal dynamics of individual districts.\\

\begin{figure}
  \centering
\includegraphics[width=15cm]{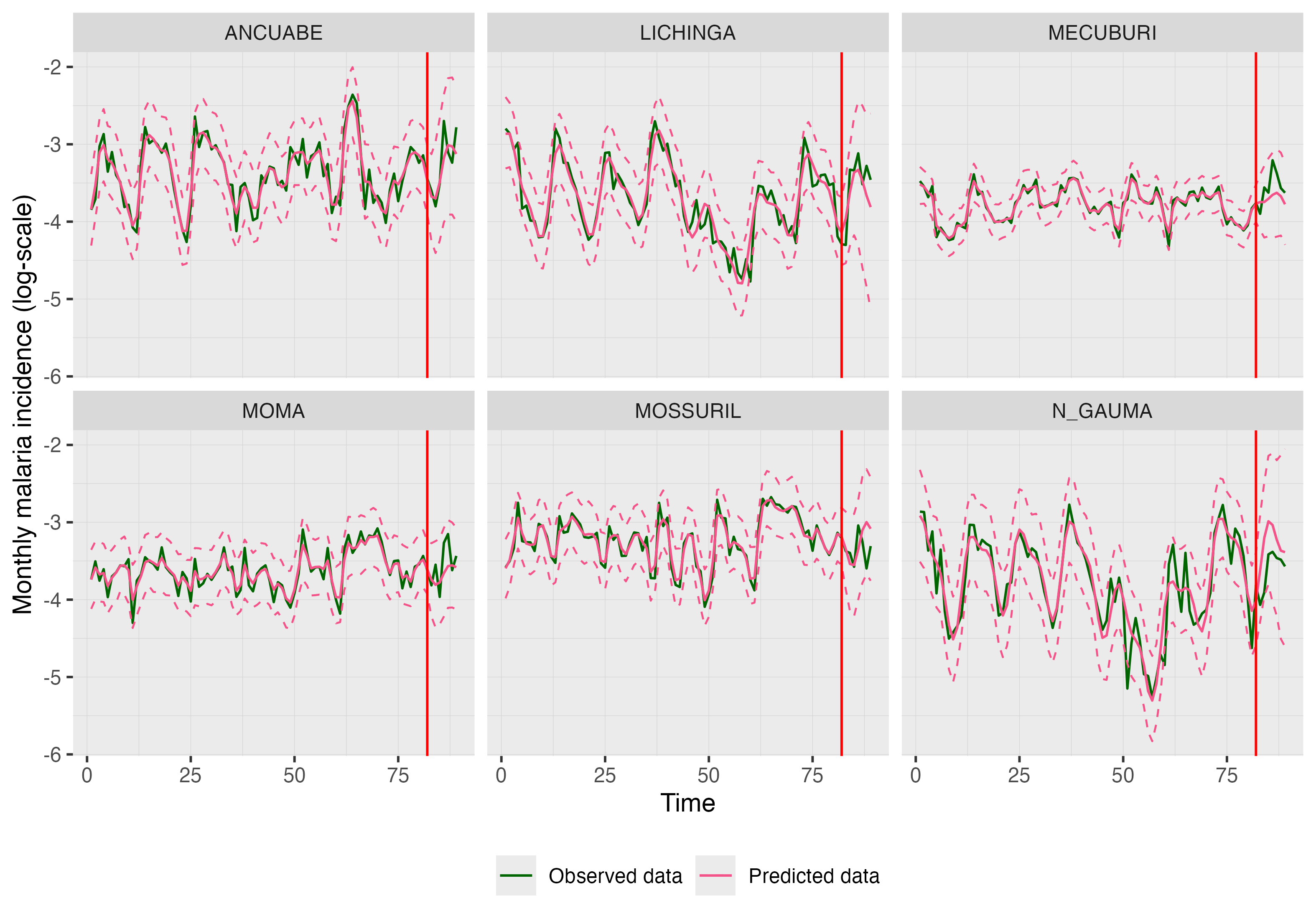}
  \caption{Model-based incidence estimates obtained from the proposed Gaussian-process time-series model for the monthly malaria incidence in six selected districts of the study area: Ancuabe, Lichinga, Mecuburi, Moma, Mossuril, N'Gauma. The results are presented on the log scale and include both point estimates and the corresponding 95\% credible intervals. The vertical red line indicates the division between the training set (before) and test set (after).}
  \label{fig:fitvsobs_malaria}
\end{figure}

The proposed spatio-temporal model provided a close fit to the observed malaria incidence during the training period, with 95\% credible intervals that closely followed the observed trends. This behavior was consistent across districts. At the test period, the model effectively captured the overall trend of the observed incidence, although predictive uncertainty increased relative to the training period. In most districts, the observed values remained within the 95\% predictive intervals. The agreement between observed and estimated trajectories can be seen in Figure~\ref{fig:fitvsobs_malaria}, which illustrates the results for six representative districts.  In some districts (e.g., Ancuabe and Lichinga), median forecasts closely matched the observed incidence. In others (e.g., Mossuril and N'Gauma), the forecasts slightly overestimated observed values, while in districts such as Moma and Mecuburi the model tended to somewhat underestimate the incidence. \\

As mentioned in Section \ref{Sect.male.performance}, four models were used as benchmarks. Model 1 follows the specification proposed by Knorr-Held \cite{knorr2000bayesian}, who introduced four types of space-time interaction structures.  We fitted these models using the \texttt{R-INLA}  \cite{INLAsoft} version 24.06.27. Among these, the type II interaction produced the best predictive fit, as it yielded the lowest Watanabe-Akaike information criterion (WAIC) and the lowest logarithmic score based on the Conditional Predictive Ordinate (CPO) score ($-\sum \log(\mathrm{CPO})$).  Details of the model selection among Knorr-Held specifications, model implementation, prior specification and posterior summaries of the main hyperparameters are provided in Appendix C. \\

To fit Models M2 and M3, we ran 50,000 MCMC iterations, with a burn-in period of 10,000 iterations and a thinning interval of 10, yielding 4,000 posterior samples for inference. The models were fitted using version 4.0 of the \texttt{R} package \texttt{CARBayesST} \cite{lee2018spatio}, adopting the package's default prior specifications. Preliminary runs indicated that convergence was achieved well before the end of the burn-in period. The retained posterior samples yielded effective sample sizes exceeding 400 for all monitored parameters, while trace plots and Geweke diagnostics indicated satisfactory convergence. Additional implementation details, prior specifications, convergence diagnostics, and posterior summaries of the main hyperparameters are provided in Appendix D.\\

Table \ref{Tab.RMSEWAIC_comp} summarizes the predictive performance of the competing models by comparing the RMSE. The RMSE was computed separately for each province by aggregating the district-level prediction errors, as well as for the complete study region, to assess the performance at different spatial scales and identify potential regional differences. The proposed GP model with correlated hyperparameters (M5) achieved the lowest RMSE in both the training and test sets.  \\ 
\begin{table}[ht]
\centering
\begin{tabular}{l cccc cccc}
\toprule
& \multicolumn{4}{c}{Training set (RMSE)} & \multicolumn{4}{c}{Test set (RMSE)}\\
\cmidrule(lr){2-5}\cmidrule(lr){6-9}
Model  &
\multicolumn{3}{c}{Province} & Total &
\multicolumn{3}{c}{Province} & Total \\
\cmidrule(lr){2-4}\cmidrule(lr){6-8}
& Cabo Delgado & Nampula & Niassa & &
Cabo Delgado & Nampula & Niassa & \\
\midrule
M1: Knorr-Held (Type II) &
2.29 & 1.11 & 1.08 & 2.77 &
2.17 & 1.35 & 2.90 & 3.86 \\

M2: Rushworth AR(1) &
2.07 & 1.06 & 1.13 & 2.59 &
2.15 & 1.35 & 2.81 & 3.79 \\

M3: Rushworth AR(2) &
2.03 & 1.04 & 1.09 & 2.53 &
2.17 & 1.38 & 2.91 & 3.89 \\

M4: Independent GP &
1.14 & 0.72 & 0.82 & 1.57 &
1.90 & \textbf{1.26} & 2.30 & 3.24 \\

M5: GP with correlated hyperparameters &
\textbf{1.03} & \textbf{0.57 }& \textbf{0.78} & \textbf{1.41} &
\textbf{1.88 }& \textbf{1.26 }& \textbf{1.95} & \textbf{2.98} \\
\bottomrule
\end{tabular}
\caption{Comparison of model performance on the training and test sets for malaria incidence in Mozambique, using the Root Mean Squared Error (RMSE), reported by province and for the complete study region.}
\label{Tab.RMSEWAIC_comp}
\end{table}

\begin{figure}[!htbp]
  \centering
  \includegraphics[width=13cm]{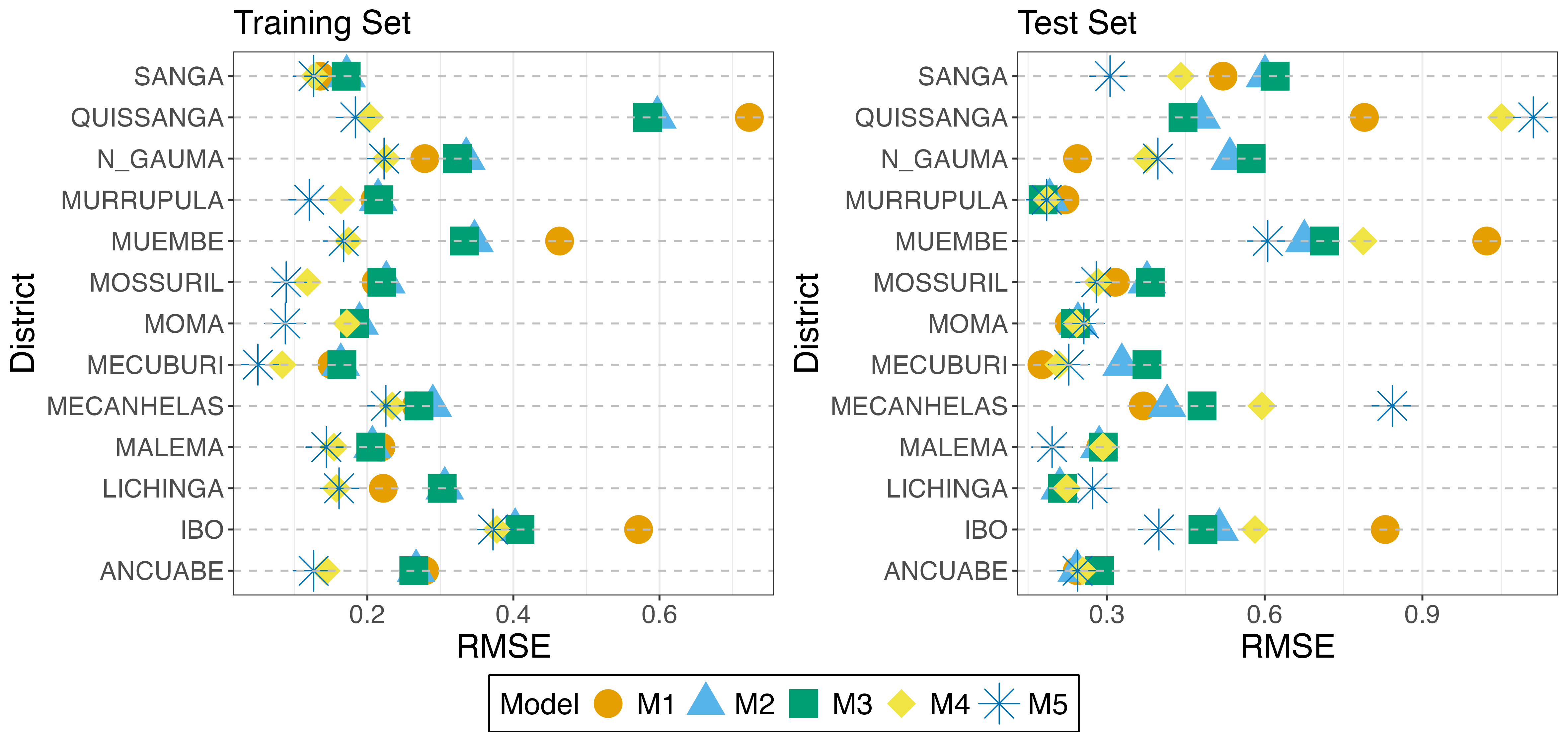}
  \caption{Comparison of model performance on the \textbf{training set} and \textbf{test set} for Malaria incidence in selected districts of the study area in Mozambique, using the Root Mean Squared Error (RMSE). M1: Knorr-Held \cite{knorr2000bayesian}, M2: first-order Rushworth et al.\cite{rushworth2014spatio}, M3: second-order Rushworth et al. \cite{rushworth2014spatio}, M4: independent Gaussian-process, M5: Gaussian-process time-series model with spatially correlated hyperparameters.} 
  \label{fig:comp_rmse_malaria}
\end{figure}

District-level RMSE values for selected districts are shown in Figure~\ref{fig:comp_rmse_malaria} . On the training set, the GP-based models (M4 \& M5) consistently achieved the lowest RMSE across most districts, indicating stronger agreement between observed and estimated incidence levels during model fitting. When performance was evaluated on the test set,  the GP-based models achieved the lowest RMSE in most districts, however, differences among models are less evident.  As expected, RMSE values are generally higher on the test set for all models, reflecting the greater difficulty of forecasting beyond the training period and indicating some degree of overfitting across approaches.\\

\begin{figure}[!htbp]
  \centering
  \includegraphics[width=13cm]{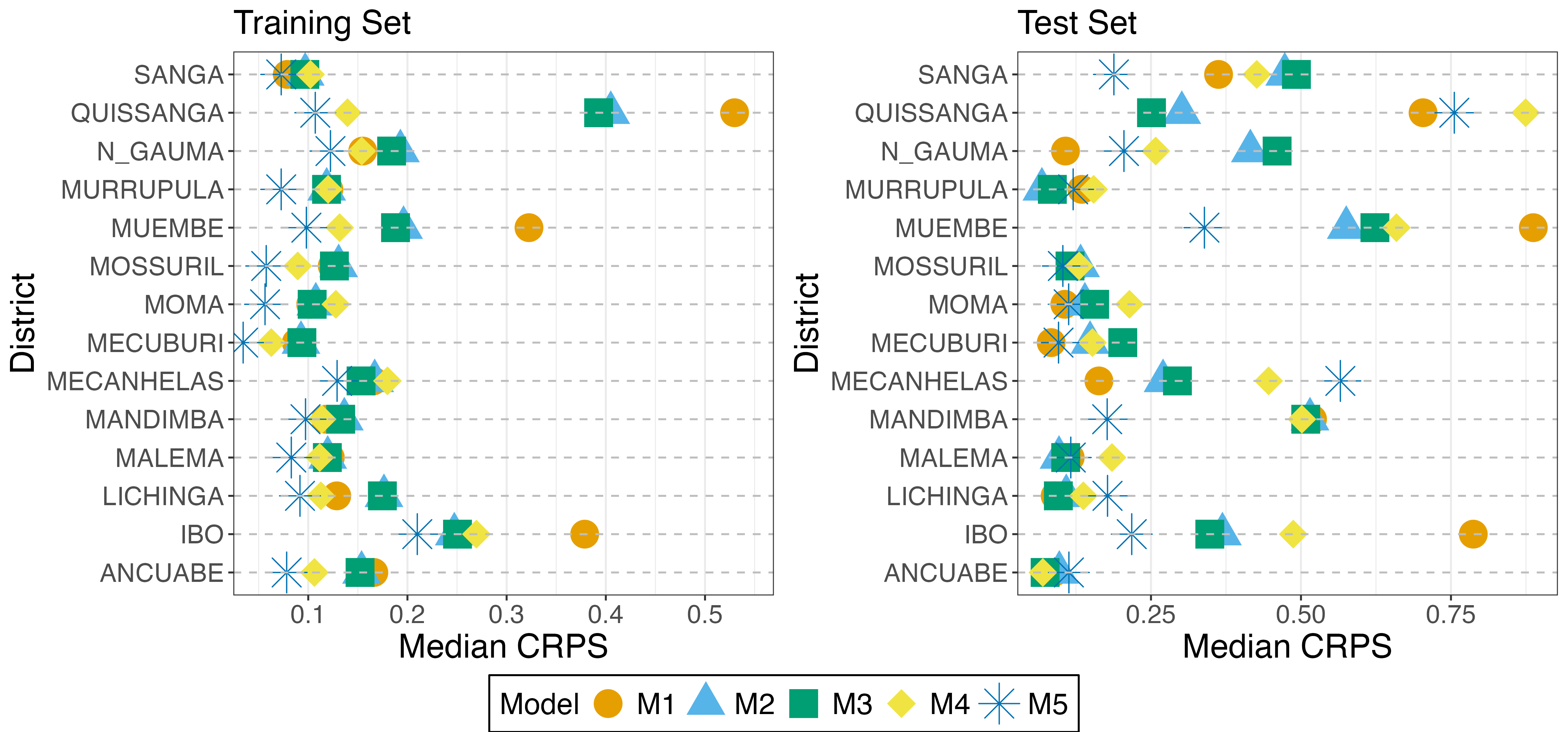}
  \caption{Comparison of model performance on the \textbf{training set} and \textbf{test set} for Malaria incidence in selected districts of the study area in Mozambique, using the Continuous Ranked Probability Score (CRPS). M1: Knorr-Held \cite{knorr2000bayesian}, M2: first-order Rushworth et al.\cite{rushworth2014spatio}, M3: second-order Rushworth et al. \cite{rushworth2014spatio}, M4: independent Gaussian-process, M5: Gaussian-process time-series model with spatially correlated hyperparameters.} 
  \label{fig:comp_crps_malaria}
\end{figure}

The district-level median CRPS values for the training and test sets are shown in Figure~\ref{fig:comp_crps_malaria}.  On the training set, the proposed model (M5) generally achieved the lowest median CRPS values across districts, although differences with the Rushworth et al. \cite{rushworth2014spatio} models (M2 and M3) and the independent GP model (M4) were often small. When evaluated on the test set, variability in CRPS increased across districts, and no single model uniformly outperformed the others. Nevertheless, the proposed model (M5) frequently remained among the best-performing models, while M2, M3, and M4 exhibited comparable probabilistic performance depending on the district. Overall, these results suggested that our proposed modeling approach maintains competitive probabilistic accuracy while providing more stable calibration across districts. Figures~\ref{fig:comp_rmse_malaria} and \ref{fig:comp_crps_malaria} display results for a subset of districts for clarity, but similar patters are observed across all districts. The full set of results can be explored through the Shiny application. \\

\begin{figure}[!htbp]
  \centering
  \includegraphics[width=13cm]{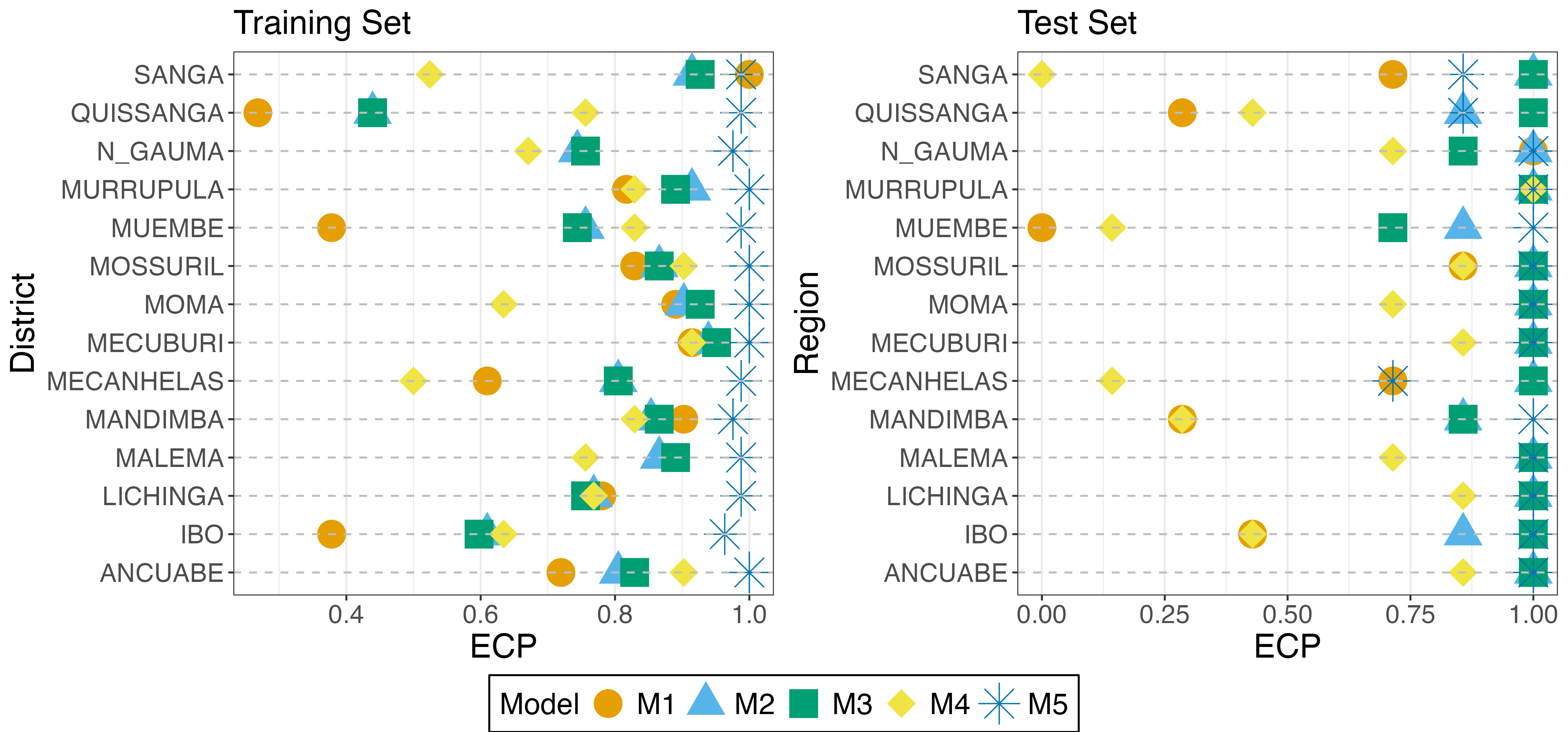}
  \caption{Comparison of model calibration on the \textbf{training set} and \textbf{test set} for Malaria incidence in selected districts of the study area in Mozambique, using the Empirical Coverage Probability (ECP). M1: Knorr-Held \cite{knorr2000bayesian}, M2: first-order Rushworth et al.\cite{rushworth2014spatio}, M3: second-order Rushworth et al. \cite{rushworth2014spatio}, M4: independent Gaussian-process, M5: Gaussian-process time-series model with spatially correlated hyperparameters.} 
  \label{fig:comp_ecp_malaria}
\end{figure}

The district-level ECP values for a representative subset of districts in the training and test sets are displayed in Figure~\ref{fig:comp_ecp_malaria}. On the training set, the proposed model (M5) consistently achieved empirical coverage probability close to one, indicating highly conservative prediction intervals. The Rushworth et al. \cite{rushworth2014spatio} (M2 \& M3) also showed high coverage, although with greater variability between districts. In contrast, the Knorr-Held model (M1) and the independent GP model (M4) frequently produced substantially lower coverage, suggesting that their predictive intervals tended to underestimate uncertainty in several districts. Similar patterns were observed on the test set. The proposed model maintained empirical coverage close to one across all districts, while M2 and M3 continued to provide relatively high and stable coverage. Overall, these results indicate that the proposed model produces the most reliable uncertainty quantification, albeit at the expense of conservative predictive intervals, whereas M1 and M4 often underestimate predictive uncertainty. It should be noted that the test set contains relatively few time points, and therefore the ECP estimates should be interpreted with caution. \\

\begin{figure}[!htbp]
  \centering
  \includegraphics[width=13cm]{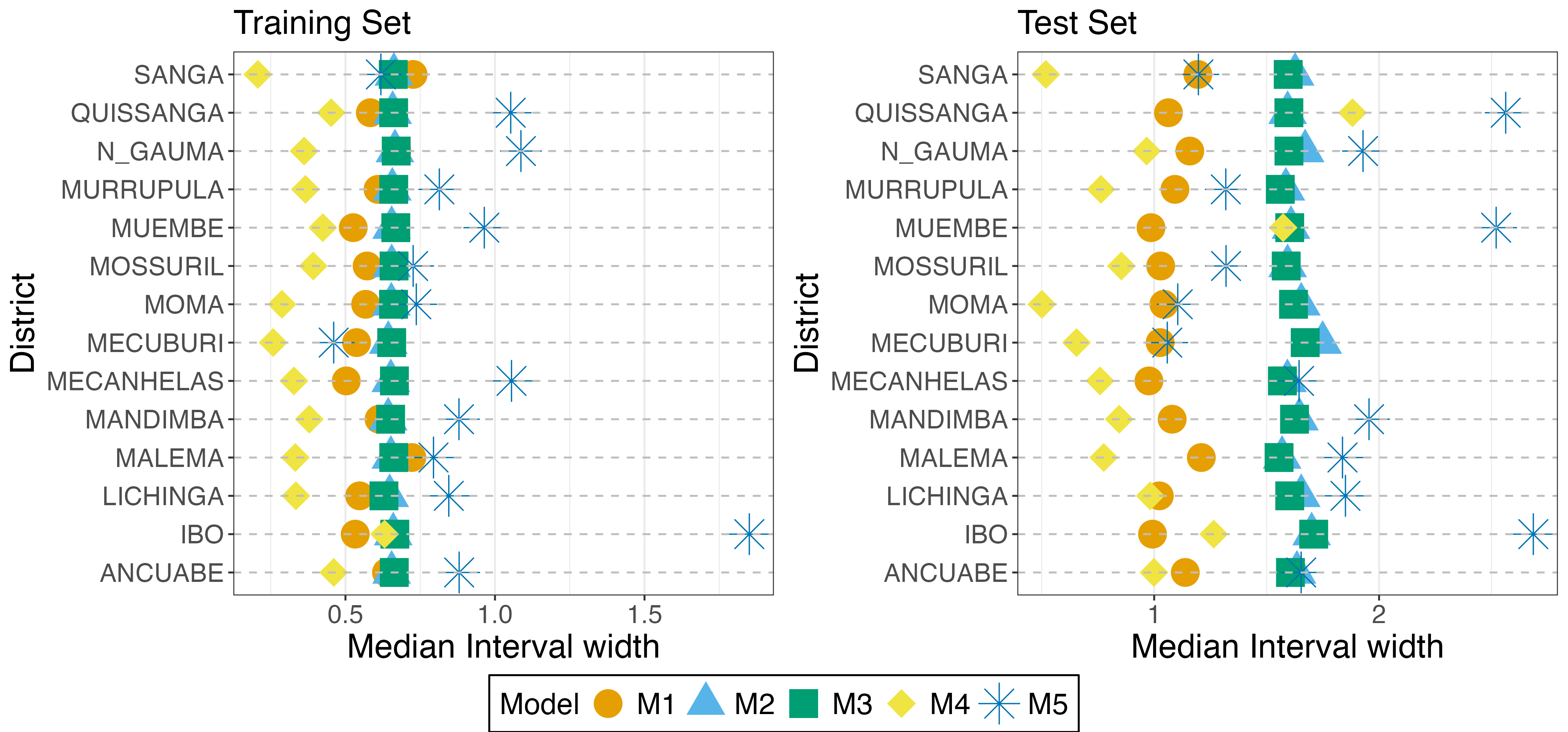}
  \caption{Comparison of model uncertainty on the \textbf{training set} and \textbf{test set} for Malaria incidence in selected districts of the study area in Mozambique, using the median credible interval width. M1: Knorr-Held \cite{knorr2000bayesian}, M2: first-order Rushworth et al.\cite{rushworth2014spatio}, M3: second-order Rushworth et al. \cite{rushworth2014spatio}, M4: independent Gaussian-process, M5: Gaussian-process time-series model with spatially correlated hyperparameters.} 
  \label{fig:comp_iw_malaria}
\end{figure}

We further compared models through the median width of the prediction intervals. The district-level interval widths for the training and test sets are displayed in Figure~\ref{fig:comp_iw_malaria}. On the training set, the proposed model (M5) generally produced the largest median interval widths across districts, reflecting a more conservative representation of predictive uncertainty during model fitting. Similar patterns were observed on the test set. The proposed model continued to generate the widest prediction intervals, while M2 and M3 produced slightly narrower but comparable interval widths. \\

When considered jointly with the ECP results, these findings highlight the trade-off between calibration and sharpness. The independent GP model (M4) achieved the narrowest prediction intervals but frequently exhibited empirical coverage below the nominal level, indicating that predictive uncertainty is underestimated. Conversely, the proposed model consistently attained empirical coverage close to one, but this is accompanied by substantially wider prediction intervals, reflecting conservative uncertainty quantification. The Rushworth et al. models (M2 and M3) provide an intermediate compromise, producing interval widths that are narrower than those of the proposed model while generally maintaining high empirical coverage. Overall, the proposed model prioritizes reliable coverage over interval sharpness, whereas M4 emphasizes sharp predictions at the expense of calibration.\\

Taken together, the predictive metrics suggest that the proposed model provides robust performance across point prediction accuracy, probabilistic forecasting, and uncertainty quantification. In terms of predictive accuracy, particularly on the test set, the proposed model performs comparably to the strongest competing spatio-temporal approaches (M2 and M3) across most districts, while consistently outperforming the Knorr-Held model (M1) in forecasting. The CRPS results indicate that the proposed model is among the best-performing models across districts, demonstrating stable probabilistic predictions without a clear loss in out-of-sample performance. Regarding uncertainty quantification, the proposed model consistently achieves empirical coverage probabilities close to one, indicating conservative predictive intervals. This conservative behavior is accompanied by wider prediction intervals than those of the competing models, reflecting a trade-off between calibration and sharpness. In contrast, the independent GP model (M4) produces considerably narrower intervals but frequently underestimates predictive uncertainty, resulting in empirical coverage below the nominal level. Overall, although all models exhibit some deterioration in performance when moving from the training to the test data, the proposed model maintained stable predictive behavior across districts, providing reliable uncertainty quantification at the expense of increased interval width and with limited evidence of severe overfitting.

\section{Discussion}
\label{Sect.discuss}

In this work, we developed a hierarchical spatio-temporal modeling framework in which temporal dynamics are modeled through Gaussian processes and spatial dependence is incorporated at the level of the hyperparameters governing temporal variability. In contrast to many traditional spatio-temporal models that impose spatial structure directly on latent risk surfaces, the proposed approach introduces spatial correlation in the parameters controlling temporal variation. This formulation provides an alternative way of representing spatio-temporal dependence, allowing neighboring regions to share information on the characteristics of their temporal dynamics rather than on the outcome process itself. Applied to malaria incidence data from Mozambique, the proposed model demonstrated competitive predictive performance with the established spatio-temporal approaches across multiple evaluation metrics. While no single model uniformly dominated across all measures, the proposed framework achieved comparable point prediction accuracy and probabilistic forecasting while providing conservative but reliable uncertainty quantification across districts. These findings suggest that modeling spatial dependence at the level of temporal variability can provide a viable and interesting alternative to traditional outcome-level spatial smoothing.\\

Our approach shares key objectives with traditional spatio-temporal models, namely smoothing estimates across space and time while quantifying uncertainty in unobserved locations. However, it differs in how spatial dependence is introduced. Classical spatio-temporal models typically extend spatial modeling frameworks to the temporal domain by imposing parametric temporal structures, such as autoregressive or random-walk processes, on latent risk surfaces. In contrast, our approach begins from a time-series perspective and then generalizes toward the spatial setting. Temporal dynamics are modeled using Gaussian processes, while spatial dependence is introduced through the hyperparameters governing temporal variability. In this way, spatial association arises through similarities in temporal dynamics across neighboring regions rather than through direct spatial smoothing of the outcome surface. \\

The idea of introducing spatial dependence through model parameters rather than directly through the outcome process is not entirely unprecedented. Similar motivations exist, for example, Rushworth et al. \cite{rushworth2014spatio} and Lee and Lawson \cite{lee2016} introduced spatial dependence through variance structures, while other approaches such as Song et al. \cite{song2020spatiotemporally} and Murakami et al. \cite{murakami2025fast} imposed spatial correlation on regression coefficients rather than on latent risk processes. These models are commonly referred to as spatio-temporally varying coefficient models \cite{song2020spatiotemporally, murakami2025fast}. The proposed model can therefore be viewed as another exploration of how spatial dependence may be introduced in the parameter space of hierarchical spatio-temporal models, here specifically through the hyperparameters governing temporal dynamics. Although these approaches share conceptual similarities with our framework, direct empirical comparisons were not included for most of the models mentioned here, except for the model proposed by Rushworth et al. \cite{rushworth2014spatio}. Many of these approaches currently lack readily available implementations or would require substantial re-implementation. Instead, we focused the empirical evaluation on widely used spatio-temporal models that are accessible through established software, allowing the proposed method to be assessed against common baseline approaches. Nevertheless, comparing alternative parameter-space formulations of spatial dependence remains an important direction for future work. \\

A key strength of the proposed framework lies in its treatment of temporal dependence. By representing temporal dynamics through Gaussian processes, the model accommodates for flexible temporal correlation structures that are not restricted to the short-memory dynamics typically assumed in autoregressive or random-walk formulations. This allows the model to capture both short- and long-range temporal dependence when supported by the data, making it well suited to disease surveillance applications where incidence may exhibit persistent temporal patterns driven by environmental, climatic, or epidemiological factors. Moreover, by introducing spatial dependence through the hyperparameters governing the temporal processes, the proposed framework departs from the conventional approach of imposing spatial structure directly on the latent outcome surface. Despite adopting a different conceptual formulation from traditional spatio-temporal models, the predictive accuracy of the proposed approach was generally comparable to that of established models across districts. These results suggest that modeling spatial dependence through the parameters governing temporal variability provides a competitive alternative to traditional outcome-level spatial smoothing while retaining the flexibility afforded by Gaussian-process representations of temporal dynamics.\\

Another strength of the proposed model is its ability to provide robust uncertainty quantification. Across both the training and test sets, the model achieved competitive CRPS values while consistently producing empirical coverage probabilities close to one across districts, indicating that its predictive intervals rarely failed to contain the observed values. This high level of coverage was accompanied by wider prediction intervals than those produced by the competing models, reflecting a conservative characterization of predictive uncertainty. Although this conservative behavior comes at the expense of interval sharpness, it avoids the substantial undercoverage observed for the independent Gaussian process model. Reliable uncertainty quantification is particularly important in disease surveillance, where prediction intervals are used to support risk assessment and public health decision-making, and conservative predictions may be preferable to overconfident forecasts. \\  

The proposed modeling framework is general in its structure and allows for flexible specifications of the mean component. In the present application, Fourier terms were included to capture large-scale seasonal patterns, while residual variation was modeled through the Gaussian-process random effects and the Gaussian noise term. Although this relatively simple specification already provided satisfactory predictive performance, incorporating additional context-relevant covariates, such as environmental drivers or population mobility, could further improve explanatory power and forecasting accuracy. For the purpose of developing and evaluating the proposed methodology, we intentionally kept the mean structure simple so that sufficient spatio-temporal variability remained in the residual component of the model.\\

There are, however, a number of limitations to our proposed modeling framework. First, the present study considered a Gaussian model applied to log-transformed incidence data. While this specification provides a convenient framework for methodological development, distributions from the exponential family would offer a more natural representation for count data. The use of the logarithmic transformation, however, is closely related to the canonical link function of a Poisson generalized linear model and yields approximately variance-stabilized and more symmetric responses. Under this perspective, the Gaussian formulation can be interpreted as a latent Gaussian representation on the link scale of the underlying non-Gaussian data-generating process. Nevertheless, extending the proposed framework to generalized linear modeling settings that directly accommodate count distributions remains an important direction for future work. \\

Second, some parameters governing the Gaussian processes, particularly the temporal variance and range parameters, exhibited relatively slower mixing, suggesting weak identifiability. Such behavior is not uncommon in hierarchical Gaussian-process models and has been documented in related literature \cite{zhang2004inconsistent, fuglstad2019constructing}, where they showed that different combinations of variance and range parameters can lead to similar covariance structures and likelihood values. In our framework, this can be partially caused by the conditional prior specification, which indirectly introduces spatial influence on the temporal range. However, these identifiability challenges did not appear to adversely affect predictive performance. A detailed discussion of model identifiability is provided in the Appendix B. Third, as in many spatio-temporal forecasting settings, the proposed model exhibited some degree of overfitting, as indicated by the expected deterioration in predictive accuracy between training and test sets. Nevertheless, it remained comparable to the degree of overfitting observed in the benchmark models and maintained stable predictive performance on unseen data.\\ 

Beyond the limitations discussed above, several options for future research arise from this work. In addition to extending the framework to alternative distributions within the exponential family and improving computational efficiency, further developments could explore richer sets of covariates, including variables related to disease intervention measures, population mobility, and environmental drivers of mosquito abundance and transmission. Incorporating such predictors could improve both interpretability and predictive performance in epidemiological applications. Furthermore, applying the model to larger spatial domains, such as malaria incidence across all provinces of Mozambique, would allow a more comprehensive assessment of its scalability and robustness. Another promising direction concerns the potential interaction between latent temporal processes and observed covariates. Because Gaussian-process components can capture complex temporal variation, part of the variability explained by the GP may overlap with effects that could alternatively be attributed to observed predictors. Investigating strategies to better distinguish between latent temporal dynamics and covariate effects therefore represents an important direction for further methodological development.\\

\section{Conclusion}
\label{Sect.conclusion}
We have presented a flexible and interpretable modeling framework for spatio-temporal areal data that combines temporal Gaussian process with spatially structured hyperparameters governing temporal variability. By introducing spatial dependence at the level of the temporal processes rather than directly on the latent outcome surface, the proposed model provides an alternative representation of spatio-temporal dependence in which neighboring regions share information about the characteristics of their temporal evolution. Applied to malaria incidence data from Mozambique, the proposed framework achieved predictive performance comparable to that of established spatio-temporal models across multiple evaluation metrics while providing robust, but conservative, uncertainty quantification. These findings demonstrate that spatial borrowing through temporal-process parameters can offer a competitive alternative to conventional outcome-level spatial smoothing without sacrificing predictive performance. More broadly, the proposed framework contributes to the growing class of hierarchical spatio-temporal models that introduce dependence through the parameter space of latent temporal processes, offering a flexible addition to the methodological toolbox for the analysis of spatio-temporal areal data.

\bmsection*{Data and code availability} 

The data used in this manuscript for the \textit{malaria data} example consist of routine malaria surveillance data collected from health centers in Mozambique. Although these data do not contain identifiable personal information, they are considered to be under the ownership of the National Malaria Control Program and are therefore not publicly available, as their use and interpretation may be sensitive at the country level. Access to the data may be requested by submitting a letter describing the proposed use and justification to the Director of the Mozambique National Malaria Control Program. 
The taylor-made algorithm and the Stan code used to implement the proposed model and reproduce the analyses presented in this study is publicly available at: \url{https://anonymous.4open.science/r/atimeseriesGP25/}. The repository includes scripts for model estimation as well as a reproducible example based on a simplified and simulated dataset consisting of 16 districts in Mozambique. 

\bmsection*{Declaration of generative artificial intelligence use} 

Generative artificial intelligence tools were used solely to assist with grammar and language editing and to enhance the visual presentation of selected figures. All scientific content, methodological decisions, statistical analyses, interpretation of the results, and final revisions were undertaken and verified by the authors. The authors take full responsibility for the content of the manuscript.

\bmsection*{Conflict of interest statement}
The authors declare that they have no conflicts of interest relevant to this work.

%\bmsection*{Author contributions}
%\bmsection*{Acknowledgments}
%TN gratefully acknowledges support from Research Foundation - Flanders (no. G0A3M24N). 
%\bmsection*{Financial disclosure}
%\bmsection*{Conflict of interest}
\bibliographystyle{vancouver}
\bibliography{wileyNJD-Vancouver.bib}
%\bmsection*{Supporting information}
\newpage

%\documentclass[vancouver,Times1COL]{WileyNJDv5}
%\documentclass[HARVARD,LATO2COL]{WileyNJDv5}
%\usepackage{float}
%\articletype{Article Type}%

% --- Supplementary material title ---
%\title{\textbf{Supplementary Material}}
%\begin{document}
\appendix
\section{Tailor-made MCMC algorithm}
\label{APp.tailor-made}

To perform Bayesian inference under the proposed hierarchical model, we developed a tailor-made Markov chain Monte Carlo (MCMC) algorithm. The primary motivation for implementing a custom sampler was twofold. First, it allowed us to investigate the computational benefits of gradient-based MCMC methods for a high-dimensional hierarchical model. Second, it provided greater flexibility to study the influence of prior specifications and the impact of estimating different sets of hyperparameters from the data. \\

For the algorithmic development, we considered the Gaussian version of the proposed model by assuming that the log-transformed monthly malaria incidence, $\log(Y_{it})$, follows a Gaussian distribution. This assumption was adopted because it yields a computationally convenient framework for deriving and studying the MCMC algorithm. The latent linear predictor is defined as, 
\begin{equation*}
\begin{aligned}
\log(Y_{it}) &= N(\eta_{it}, \tau^2_i),\\
\eta_{it} &= \beta_{0i}
+\beta_{1i}\sin\!\left(\frac{2\pi t}{12}\right)
+\beta_{2i}\sin\!\left(\frac{2\pi t}{6}\right)\\
&\quad+\beta_{3i}\cos\!\left(\frac{2\pi t}{12}\right)
+\beta_{4i}\cos\!\left(\frac{2\pi t}{6}\right)
+Z_{it},
\end{aligned}
\label{mod.malaria}
\end{equation*}
where $\mathbf{Z}_i=(Z_{i1},\ldots,Z_{iT})$ follows a Gaussian process with a Matérn covariance function. We opted for the Matérn covariance function since it is typically used in spatial and spatio-temporal modeling. \\

Inference targets the joint posterior distribution of the district-specific parameters
$$
\boldsymbol{\Theta}=(\theta_1,\ldots,\theta_I),\qquad
\theta_i=\left(\boldsymbol{\beta}_i,\log(\sigma_i^2),\log(\phi_i),\log(\nu_i^2),\mathbf{Z}_i\right),
$$
where $\nu_i^2=\tau_i^2/\sigma_i^2$ denotes the signal-to-noise ratio. We parameterized the model in terms of $\nu_i^2$ rather than the nugget variance $\tau_i^2$, following the recommendation of Diggle and Giorgi \cite{diggle2019model}. This reparameterization simplifies several gradient expressions and improves the efficiency of the gradient-based updates described below. \\

Since the posterior distribution $p(\boldsymbol{\Theta}\mid\mathbf{y})$ is not available in closed form, inference is performed using MCMC. Throughout this appendix, $k$ denotes the MCMC iteration, superscript $(k)$ indicates the current state of the chain, and proposed values are denoted by a tilde. The sampler updated the parameters in blocks by combining Metropolis-adjusted Langevin algorithm (MALA)  \cite{mala, roberts1996exponential}, Metropolis--Hastings (MH) \cite{metropolis1953, hastings1970}, and Gibbs updates \cite{gelfand1990}. MALA was employed for the log-variance and log-range parameters because gradient information substantially improves exploration of the posterior distribution. The log signal-to-noise ratio was updated using a Metropolis--Hastings step, while Gibbs sampling was used for the regression coefficients and latent Gaussian process effects, whose full conditional distributions are available. The following sections describe the form of priors and hyperpriors, and each update in detail.

\subsection{Prior specification}
This section describes the prior and hyperprior distributions adopted  for the proposed model. We restricted the variance, the range and the signal-to-noise ratio parameters to be positive, therefore, they were modeled on the logarithmic scale. This constrain allow prior distributions to be supported on the entire real line. \\ 

Specifically, we assumed that the vector of log-variances $\log(\boldsymbol{\sigma^2}) = (\log(\sigma^2_1), \ldots, \log(\sigma^2_I))$ follows a conditional autoregressive (CAR) prior of the Leroux form,
\begin{equation*}
\log(\boldsymbol{\sigma^2}) \sim \mathcal{N}\big(\mathbf{0}, \mathbf{Q}^{-1}\big),
\end{equation*}
where
$$
\mathbf{Q} = \delta^2 \big[(1-\alpha)\mathbf{I} + \alpha(\mathbf{K} - \mathbf{W})\big].
$$
Here, the matrix $\mathbf{W}$ is the Queen contiguity adjacency matrix, $\mathbf{K}$ is the diagonal matrix of neighborhood counts and $\delta^2$ controls the overall precision and was assigned a Gamma(1, 0.001) hyperprior. The spatial dependence was fixed at  $\alpha = 0.9$, based on preliminary Besag-York-Mollié (BYM) \cite{besag1991bayesian} fits to the region-specific variance estimates obtained from independent Gaussian-process models. Preliminary experiments indicated that estimating $\alpha$ substantially increased the computational cost while having little influence on posterior inference.  \\

The log-range parameter was specified conditionally on the log-variance through,
\begin{equation*}
\log(\phi_i)\mid \log(\sigma_i^2) \sim \mathcal{N}\big(\mu_{\phi_i},\,\sigma^2_{\phi_i}\big),
\end{equation*}
where
\begin{equation*}
\mu_{\phi_i} = \widehat{\log(\phi_i)}_{\text{MLE}} +
\frac{\widehat{\operatorname{Cov}}(\log(\phi_i), \log(\sigma_i^2))}
{\widehat{\operatorname{Var}}(\log(\sigma_i^2))}
\big(\log(\sigma_i^2) - \widehat{\log(\sigma_i^2)}_{\text{MLE}}\big),
\end{equation*}
and
\begin{equation*}
\sigma^2_{\phi_i} =
\widehat{\operatorname{Var}}(\log(\phi_i)) -
\frac{\widehat{\operatorname{Cov}}(\log(\phi_i), \log(\sigma_i^2))^2}
{\widehat{\operatorname{Var}}(\log(\sigma_i^2))}.
\end{equation*}
This conditional specification preserves the empirical association between the variance and range parameters observed in district-specific Gaussian-process fits. The required moments were estimated by fitting independent Gaussian-process models separately to each district using maximum likelihood. Using preliminary maximum-likelihood estimates to inform prior distributions is a common strategy in hierarchical modeling, as it allows prior information to be centered on plausible parameter values, facilitating exploration of the parameter space and improving computational efficiency.\\

The signal-to-noise ratio, $\nu_i^2 = \tau_i^2/\sigma_i^2$ was assigned the prior:
$$
\log(\nu_i^2) \sim \mathcal{N}\big(\widehat{\log(\nu_i^2)}_{\text{MLE}},\,\widehat{\operatorname{Var}}(\log(\nu_i^2))\big),
$$
again using empirical moments obtained from the region-specific Gaussian process models. Parameterizing the model through the signal-to-noise ratio, rather than directly through the nugget variance $\tau_i^2$, simplified several gradient expressions and improved the efficiency of the gradient-based updates. \\

Finally, regression coefficients were assigned Gaussian priors,
$$
\boldsymbol{\beta}_i \sim \mathcal{N}_{p+1}(0, 10^5 I_{p+1}),
$$
where $p$ denotes the number of covariates. \\

The empirical calibration adopted for the covariance parameters was motivated by the objectives of developing a tailor-made MCMC algorithm. First, centering the priors on plausible parameter values substantially improved mixing of the variance-related parameters. Second, fixing selected hyperparameters and using empirically calibrated priors made it possible to assess the influence of these modeling choices on posterior inference and to identify which quantities could instead be estimated directly from the data. These observations ultimately motivated the implementation of the proposed framework in more general Bayesian software, where efficient Hamiltonian Monte Carlo algorithms make it feasible to estimate a larger set of hyperparameters while maintaining good computational performance.

\subsection{Sampler overview}

The sampler updates parameters in blocks by combining MALA, Metropolis–Hastings, and Gibbs steps. The spatially structured log-variance parameters were updated jointly because their CAR prior induces dependence across districts. Conditional on the updated log-variance vector, the remaining parameters were updated district by district. The resulting sampler can be summarized as follows:

\begin{algorithm}
\caption{Tailor-made MCMC sampler}
\begin{algorithmic}[1]
\For{$k=0,\ldots,K-1$}
\State Update $\boldsymbol{\ell}_\sigma^{(k+1)}$ jointly using a multivariate MALA step.
\For{$i=1,\ldots,I$}
\State Update $\log(\phi_i)^{(k+1)}$ using a univariate MALA step.
\State Update $\log(\nu_i^2)^{(k+1)}$ using a random-walk Metropolis-Hastings step.
\State Sample $\boldsymbol{\beta}_i^{(k+1)}$ from its full conditional Gaussian distribution using Gibbs sampler.
\State Sample $\boldsymbol{Z}_i^{(k+1)}$ from its full conditional Gaussian distribution using Gibbs sampler.
\EndFor
\State Adapt MALA proposal step sizes during burn-in.
\EndFor
\end{algorithmic}
\end{algorithm}

\subsubsection{Step 1:  Update for log-variances}
The vector $\boldsymbol{\ell}_\sigma=\big(\log(\sigma_1^2),\ldots,\log(\sigma_I^2)\big)^\top$ was updated jointly using a multivariate MALA step with isotropic proposal covariance. At iteration $k$, the MALA step proposed: 
$$
\widetilde{\boldsymbol{\ell}}_\sigma =
\boldsymbol{\ell}_\sigma^{(k)} +
\frac{(\epsilon_\sigma^{(k)})^2}{2}\,\mathbf{g}_\sigma^{(k)} +
\epsilon_\sigma^{(k)}\,\mathbf{z},
\qquad \mathbf{z}\sim\mathcal{N}(\mathbf{0},\mathbf{I}_I).
$$ where $\epsilon_\sigma^{(k)}$ denotes the current step size, $
\mathbf{g}_\sigma^{(k)}=\nabla_{\boldsymbol{\ell}_\sigma}\log p\!\left(\boldsymbol{\ell}_\sigma^{(k)} \mid \text{rest}\right)$ denotes the gradient of the log-posterior with respect to $l_{\sigma}$.  \\
The proposal density is therefore $\mathcal{N}\!\big(\boldsymbol{\ell}_\sigma^{(k)}+\tfrac{(\epsilon_\sigma^{(k)})^2}{2}\mathbf{g}_\sigma^{(k)},\, (\epsilon_\sigma^{(k)})^2\mathbf{I}\big)$. We accepted the proposal vector $\widetilde{\boldsymbol{\ell}}_\sigma$ with probability:
$$
\alpha_\sigma = \min\left\{
1,\;
\exp\Big[
\log \pi(\widetilde{\boldsymbol{\ell}}_\sigma)-\log \pi(\boldsymbol{\ell}_\sigma^{(k)})
+\log q(\boldsymbol{\ell}_\sigma^{(k)}\mid \widetilde{\boldsymbol{\ell}}_\sigma)
-\log q(\widetilde{\boldsymbol{\ell}}_\sigma\mid \boldsymbol{\ell}_\sigma^{(k)})
\Big]
\right\},
$$
where $\pi(\cdot)$ denotes the target density of the block and $q(\cdot\mid\cdot)$ the proposal density.

\subsubsection{Adaptive step-size tuning.}
We tuned $\epsilon_\sigma^{(k)}$ using a Robbins-Monro-type \cite{robbins1951stochastic} update targeting the optimal MALA acceptance rate of $0.57$:
$$
\epsilon_\sigma^{(k+1)}
=
\left[
H\left(
\sqrt{\epsilon_\sigma^{(k)}}+\frac{1}{k+1}\big(\widehat{\alpha}_\sigma^{(k)}-0.57\big)
\right)
\right]^2,
$$
where $H(\cdot)$ truncates values to a prespecified interval $[10^{-4},10^{4}]$, and $\widehat{\alpha}_\sigma^{(k)}=\min\{1,\exp(\log\alpha_\sigma^{(k)})\}$ denotes the acceptance probability used in the code.

\subsubsection{Step 2: Area-specific updates given $\boldsymbol{\ell}_\sigma^{(k+1)}$}
Conditional on the updated $\boldsymbol{\ell}_\sigma^{(k+1)}$, the remaining parameters were updated independently for each district. For district $i$, the updates are performed in the following order:
\[
\log(\phi_i)\;\rightarrow\;\log(\nu_i^2)\;\rightarrow\;\boldsymbol{\beta}_i\;\rightarrow\;\boldsymbol{Z}_i.
\]
\\
\textbf{(a) Update for log-range parameter, $\log(\phi_i)$.} \\
The log-range parameter was updated using an univariate MALA proposal, as follows
$$
\widetilde{\log(\phi_i)} =
\log(\phi_i)^{(k)} +
\frac{(\epsilon_{\phi,i}^{(k)})^2}{2}\, g_{\phi,i}^{(k)} +
\epsilon_{\phi,i}^{(k)}\,z,\qquad z\sim\mathcal{N}(0,1),
$$
with $\epsilon_{\phi,i}^{(k)}$ the step-size and $g_{\phi,i}^{(k)}$ the gradient of the log-posterior with respect to $\log(\phi_i)$. We accepted the proposal value with the standard MALA acceptance probability using forward/backward Gaussian proposal densities. We tuned $\epsilon_{\phi,i}^{(k)}$ using the same truncated Robbins-Monro update targeting $0.57$.\\

\textbf{(b) Update for log-signal-to-noise ratio parameter, $\log(\nu_i^2)$.} \\
We updated $\log(\nu_i^2)$ using a random-walk MH step:
$$
\widetilde{\log(\nu_i^2)} \sim \mathcal{N}\!\left(\log(\nu_i^2)^{(k)},\, s_{\nu,i}^2\right),
$$
where $s_{\nu,i}$ was taken from preliminary region-specific GP fits. Because the proposal distribution was symmetric, the MH acceptance probability reduced to
$$
\alpha_{\nu,i}
=
\min\left\{
1,\;
\exp\Big[
\log \pi(\widetilde{\log(\nu_i^2)})-\log \pi(\log(\nu_i^2)^{(k)})
\Big]
\right\},
$$
where $\pi(\cdot)$ denotes the target density for this block (prior for $\log(\nu_i^2)$ plus the Gaussian likelihood).\\

\textbf{(c)  Update for regression coefficients, $\boldsymbol{\beta}_i$.} \\
Conditional on $(\boldsymbol{Z}_i,\log(\nu_i^2),\log(\sigma_i^2))$, the full conditional posterior distribution of $\boldsymbol{\beta}_i$ is a multivariate Gaussian.
Let $\tau_i^2=\exp\{\log(\nu_i^2)+\log(\sigma_i^2)\}$. Define
$$
\mathbf{P}_{\beta,i} =
\frac{1}{\tau_i^2}\mathbf{X}_i^\top \mathbf{X}_i + \lambda_\beta \mathbf{S},
\qquad
\mathbf{m}_{\beta,i} =
\mathbf{P}_{\beta,i}^{-1}\left(\frac{1}{\tau_i^2}\mathbf{X}_i^\top(\mathbf{y}_i-\boldsymbol{Z}_i)\right).
$$
Then
$$
\boldsymbol{\beta}_i \mid \text{rest} \sim \mathcal{N}\!\left(\mathbf{m}_{\beta,i},\,\mathbf{P}_{\beta,i}^{-1}\right).
$$ \\

\textbf{(d) Update for random effects, $\boldsymbol{Z}_i$.} \\
Conditional on $(\boldsymbol{\beta}_i,\sigma_i^2,\phi_i,\tau_i^2)$, the full conditional posterior distribution of $\boldsymbol{Z}_i$ is also Gaussian. Let $\mathbf{R}_i=\mathbf{R}(\phi_i)+\epsilon\mathbf{I}_T$ and note that $\boldsymbol{Z}_i\sim \mathcal{N}(\mathbf{0},\sigma_i^2\mathbf{R}_i)$ a priori. The full conditional precision is
$$
\mathbf{P}_{Z,i}
=
\frac{1}{\tau_i^2}\mathbf{I}_T + \frac{1}{\sigma_i^2}\mathbf{R}_i^{-1},
$$
and the full conditional mean is
$$
\mathbf{m}_{Z,i}
=
\mathbf{P}_{Z,i}^{-1}\left(\frac{1}{\tau_i^2}(\mathbf{y}_i-\mathbf{X}_i\boldsymbol{\beta}_i)\right).
$$
Therefore,
$$
\boldsymbol{Z}_i \mid \text{rest}\sim \mathcal{N}\!\left(\mathbf{m}_{Z,i},\,\mathbf{P}_{Z,i}^{-1}\right).
$$

\section{Discussion of model identifiability}
\label{App.identifiability}

In the proposed model, potential identifiability concerns arose among the parameters governing the Gaussian-process and observation-error components within each area-specific time series, namely the marginal variance parameter $\log(\sigma_i^2)$, the temporal range parameter $\log(\phi_i)$, and the nugget variance parameter $\log(\tau_i^2)$. As is well known for Gaussian-process models, practical identifiability issues may arise because different combinations of process variance, temporal range, and observation-error variance can produce very similar covariance structures in the observed data. This appendix investigates the extent to which this phenomenon is present in the proposed model using three complementary diagnostics.

\subsection{Posterior dependence among covariance parameters}

A first indication of weak identifiability is strong posterior dependence between covariance parameters. Table~\ref{Tab.corr} summarizes the posterior correlations for a representative subset of districts. The strongest dependence is consistently observed between $\log(\sigma_i^2)$ and $\log(\phi_i)$, whereas the correlations involving the nugget effect parameter are generally much weaker. \\

Strong posterior correlation does not imply non-identifiability by itself. However, it indicates that the likelihood contains limited information to distinguish these parameters individually, so that changes in one parameter can often be compensated by changes in another while producing a similar likelihood.

\begin{table}[h]
\centering
\begin{tabular}{c|ccccccccc}
\hline
District & 1 & 2 & 3 & 4 & 5 & 6 & 7 & 8 & 9\\
\hline
$corr(\log(\sigma_i^2),\log(\phi_i))$
&0.91&0.77&0.92&0.82&0.95&0.35&0.82&0.21&0.75\\

$corr(\log(\sigma_i^2),\log(\nu_i^2))$
&0.10&0.22&0.46&0.30&0.39&-0.09&0.30&-0.33&0.18\\

$corr(\log(\nu_i^2),\log(\phi_i))$
&0.14&0.58&0.57&0.49&0.48&0.65&0.51&0.53&0.41\\
\hline
\end{tabular}
\caption{Posterior correlations among covariance hyperparameters for representative districts.}
\label{Tab.corr}
\end{table}

\subsection{Prior and posterior distributions}

\begin{figure}[h]
\centering
\includegraphics[width=13cm]{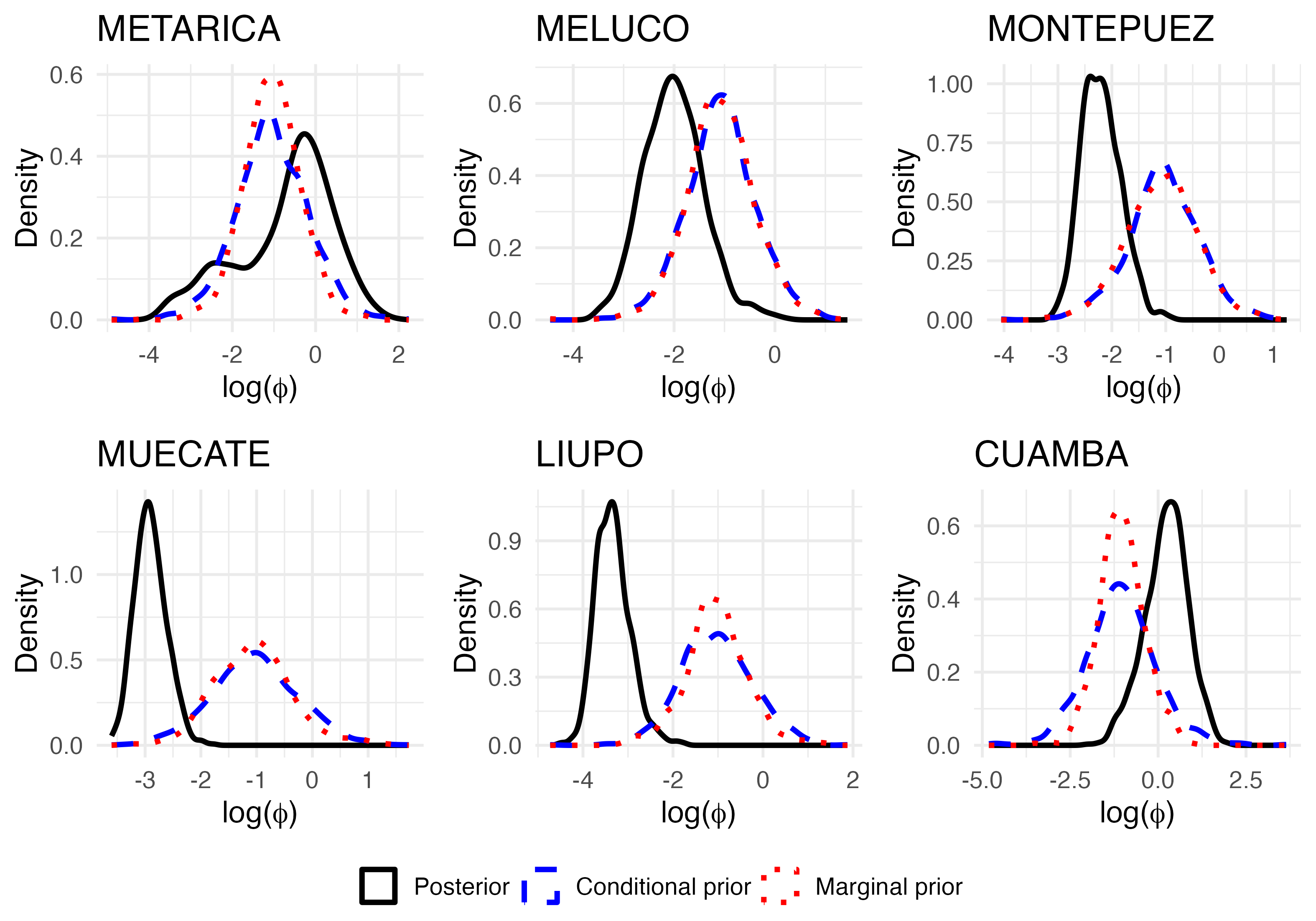}
\caption{Prior (blue and red) and posterior (black) distributions of the temporal range parameter $\log(\phi_i)$ for representative districts.}
\label{fig.post_vs_prior.phi}
\end{figure}

\begin{figure}[h]
\centering
\includegraphics[width=13cm]{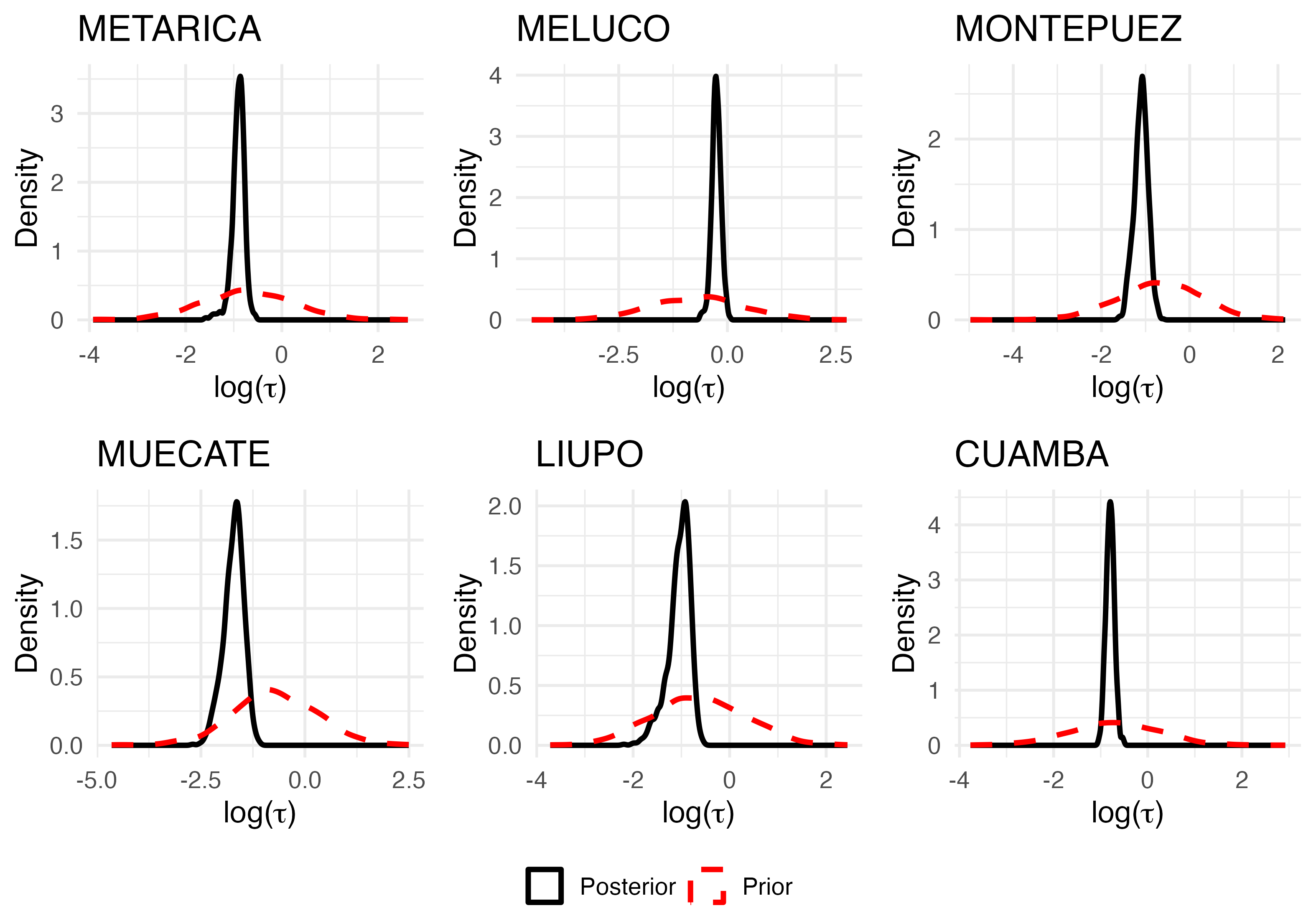}
\caption{Prior (red) and posterior (black) distributions of the signal-to-noise ratio $\log(\nu_i^2)$ for representative districts.}
\label{fig.post_vs_prior.nu}
\end{figure}

Posterior correlation alone is insufficient to assess identifiability because highly correlated parameters may still be well informed by the data. We therefore compared the prior and posterior distributions of the covariance parameters. Figure~\ref{fig.post_vs_prior.phi} compares the prior and posterior distributions of the temporal range parameter for several representative districts. Since the prior for $\log(\phi_i)$ is specified conditionally on $\log(\sigma_i^2)$, both the marginal and conditional prior distributions are shown. Figure~\ref{fig.post_vs_prior.nu} presents the corresponding comparison for the signal-to-noise ratio. For both parameters, $\log(\phi_i)$ and $\log(\tau_i^2)$, the posterior distributions are substantially more concentrated than the corresponding priors and are shifted away from the prior centers. These changes indicate that the observed data provide meaningful information about the covariance parameters despite the posterior dependence observed in Table~\ref{Tab.corr}. \\

\subsection{Joint posterior geometry}

\begin{figure}[h]
\centering
\includegraphics[width=13cm]{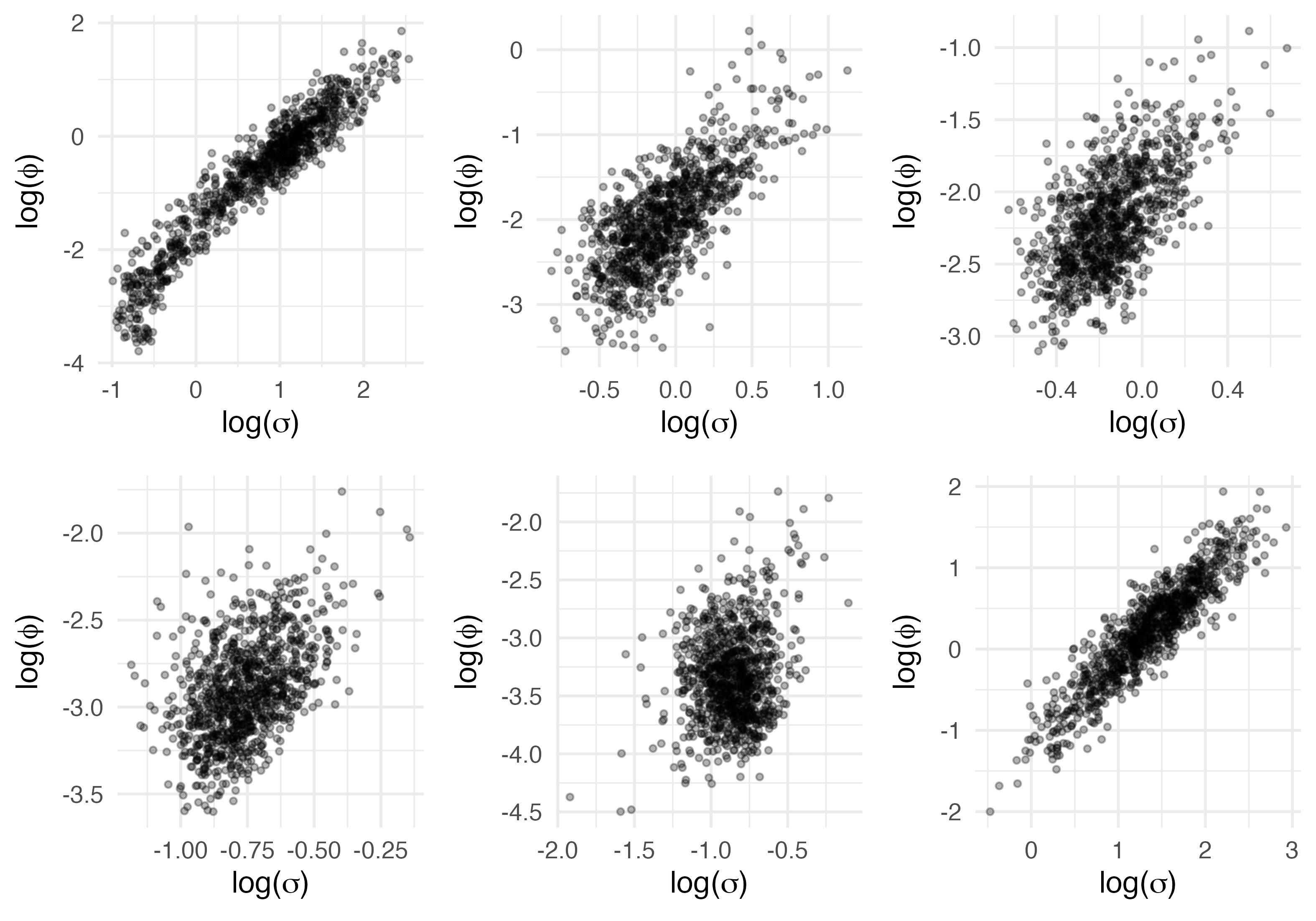}
\caption{Joint posterior distributions of $\log(\sigma_i^2)$ and $\log(\phi_i)$ for representative districts.}
\label{fig.cloud.param}
\end{figure}

To further investigate the dependence between covariance parameters, Figure~\ref{fig.cloud.param} displays posterior samples of $\log(\sigma_i^2)$ and $\log(\phi_i)$ for several representative districts. The elongated posterior clouds illustrate the strong dependence between the variance and range parameters. Similar covariance structures can be obtained by combining larger marginal variances with shorter temporal ranges, or conversely smaller marginal variances with longer temporal ranges. Consequently, the likelihood provides limited information to distinguish these parameters individually. \\

This phenomenon is well established in the Gaussian-process literature. For Matérn covariance functions, it has been shown that variance and range parameters cannot generally be estimated independently from a single realization of the process under common asymptotic regimes, and only particular combinations of these parameters are consistently estimable \cite{zhang2004inconsistent,fuglstad2019constructing}. The posterior dependence observed in Figure~\ref{fig.cloud.param} is therefore consistent with the theoretical properties of Gaussian-process covariance models rather than a deficiency of the proposed framework.\\

Despite the weak practical identifiability of the covariance hyperparameters, the latent temporal effects and predictive quantities remained stable throughout the analysis. This observation is reflected in the predictive results presented in the main text, where the proposed model exhibited competitive point prediction accuracy together with well-calibrated predictive distributions. Such behavior is consistent with previous findings in the Gaussian-process literature, where reliable prediction can often be achieved even when individual covariance parameters are only weakly identifiable.

\section{Implementation and comparison of the Knorr--Held models}

To provide a benchmark for the proposed methodology, we fitted the complete family of spatio-temporal interaction models proposed by Knorr--Held \cite{knorr2000bayesian} using the \texttt{R-INLA} package. Specifically, we considered the baseline model without interaction together with the four interaction structures described in \cite{knorr2000bayesian}: Type I (unstructured space and unstructured time), Type II (structured space and unstructured time), Type III (unstructured space and structured time), and Type IV (structured space and structured time).  All models were implemented using the BYM2 parameterization for the spatial random effect together with the default sum-to-zero constraints provided by INLA. Penalized complexity (PC) priors were adopted for the precision and mixing parameters \cite{simpson2017penalising}. These priors provided weakly informative regularization by shrinking toward simpler spatial structures while improving parameter interpretability and reducing overfitting. \\

Model comparison was performed using the Watanabe--Akaike Information Criterion (WAIC) and the logarithmic score based on the Conditional Predictive Ordinate (CPO). The resulting values are presented in Table~\ref{tab:waic_knorr_held}. Among the fitted models, the Type II interaction model achieved the lowest  WAIC and logarithmic CPO score and was therefore selected as the Knorr--Held benchmark for the predictive comparison presented in the main text. Although the Type IV interaction occasionally produced competitive predictive scores, repeated fits revealed substantial numerical instability. \\

\begin{table}[!htbp]
\centering
\begin{tabular}{lccccc}
\hline
Criterion &
Baseline &
Type I &
Type II &
Type III &
Type IV \\
\hline
WAIC &
5002 &
4635 &
4797 &
5056 &
-- \\
$-\sum\log(\mathrm{CPO})$ &
2503 &
2491 &
2390 &
2530&
-- \\
\hline
\end{tabular}
\caption{Comparison of the different Knorr-Held \cite{knorr2000bayesian} model specifications using the WAIC and the logarithmic score based on the Conditional Predictive Ordinate (CPO). Interaction type I: unstructured spatial and unstructured temporal effects interaction. Interaction type II: structured spatial and unstructured temporal effects interaction. Interaction type III: unstructured spatial and structured temporal effects interaction. Interaction type IV: structured spatial and structured temporal effects interaction.}
\label{tab:waic_knorr_held}
\end{table}

To investigate this inestability, the Type IV model was fitted repeatedly using identical model specifications. Across independent runs, the WAIC varied considerably, ranging approximately from $-23$ to $10$, suggesting convergence toward different posterior solutions despite identical initialization and prior specification.  We examined the posterior summaries and they indicated that the instability primarily affected the precision parameter associated with the BYM2 spatial effect together with the precision of the structured spatio-temporal interaction. In contrast, the remaining hyperparameters were relatively stable across runs. Consequently, the Type IV specification was not considered further in the predictive comparison. Table~\ref{tab:unstable_fit4} illustrates the posterior means obtained from four independent fits of the Type IV interaction model.
\begin{table}[!htbp]
\centering
\begin{tabular}{lcccc}
\hline
Hyperparameter &
Fit 1 &
Fit 2 &
Fit 3 &
Fit 4 \\
\hline
Precision (Gaussian) & 67.3 & 9.78 & 2.67 & 26.0\\
Precision (BYM2) & 5219.81 & 1339.00 & 7.78 & 2071.00\\
BYM2 mixing parameter ($\phi$) & 0.33 & 0.35 & 0.31 & 0.37\\
Precision (RW1) & 239.18 & 230.00 & 239.00 & 254.00\\
Precision (IID time) & 121.58 & 122.00 & 121.00 & 129.00\\
Precision (Space--time interaction) & 70.00 & 20.00 & 70.00 & 12.00\\
\hline
\end{tabular}
\caption{Posterior means of the hyperparameters obtained from four independent fits of the Type IV Knorr--Held interaction model. }
\label{tab:unstable_fit4}
\end{table}

Posterior summaries for the selected Type II interaction model are reported in Table \ref{tab:kh2_hyperparameters}.
\begin{table}[!htbp]
\centering
\begin{tabular}{lrrrrrr}
\hline
Hyperparameter &
Mean &
SD &
2.5\%&
Median &
97.5\% &
Mode\\
\hline
Precision (Gaussian) &
7.25 &
0.19 &
6.89 &
7.24 &
7.65 &
7.21\\

Precision (BYM2) &
915.42 &
12396.0 &
0.83 &
49.34 &
5588.78 &
0.81\\

BYM2 mixing parameter ($\phi$) &
0.39 &
0.30 &
0.03 &
0.32 &
0.97 &
0.06\\

Precision (RW1) &
239.85 &
195.96 &
42.28 &
186.06 &
757.50 &
108.13\\

Precision (IID time) &
116.50 &
55.80 &
44.12 &
104.73 &
258.27 &
84.99\\

Precision (Space--time interaction) &
25.83 &
3.46 &
20.09 &
25.47 &
33.66 &
24.51\\
\hline
\end{tabular}
\caption{Posterior summaries of the hyperparameters for the selected Type II Knorr--Held interaction model.}
\label{tab:kh2_hyperparameters}
\end{table}

\section{Implementation and results of the Rushworth \textit{et al.} models}
\label{App:rushworth_model}

To benchmark the proposed methodology against alternative Bayesian spatio-temporal approaches, we implemented the two conditional autoregressive models proposed by Rushworth \textit{et al.} \cite{rushworth2014spatio}. The first model assumes a first-order autoregressive process (AR) for the temporal dependence (CARAR(1)), whereas the second extends the temporal process to second-order AR specification (CARAR(2)). Both models were fitted using Markov chain Monte Carlo, following the implementation described in the original publication. \\

Preliminary runs indicated that convergence was achieved well before 10,000 iterations. Consequently, inference was based on a burn-in period of 10,000 iterations followed by 40,000 sampling iterations, retaining every tenth iteration to reduce autocorrelation. This resulted in 4,000 posterior samples for inference. Convergence was assessed using effective sample sizes (ESS) and Geweke diagnostics. Across both models, all monitored parameters exhibited ESS values exceeding 400 and Geweke statistics close to zero, indicating satisfactory mixing and convergence of the MCMC chains. \\

Posterior summaries and convergence diagnostics for the CARAR(1) model are reported in Table~\ref{tab:carar1_diagnostics}. The posterior distribution suggests strong spatial dependence ($\rho_S$ close to one), whereas the temporal autocorrelation parameter ($\rho_T$) is centered close to zero, indicating relatively weak residual temporal dependence after accounting for the seasonal structure included in the mean model.

\begin{table}[!htbp]
\centering
\caption{Posterior summaries and MCMC diagnostics for the first-order autoregressive Rushworth et al. \cite{rushworth2014spatio} model.}
\label{tab:carar1_diagnostics}
\begin{tabular}{lrrrrrrr}
\toprule
Parameter & Mean & 2.5\% & 97.5\% & $n_{\mathrm{sample}}$ & Acceptance (\%) & ESS & Geweke \\
\midrule
$\nu^{2}$     & 0.1288 & 0.1212 & 0.1363 & 4000 & 100.0 & 1477.8 & 0.3 \\
$\tau^{2}$    & 0.0686 & 0.0553 & 0.0840 & 4000 & 100.0 & 544.8  & -0.8 \\
$\rho_{S}$    & 0.9818 & 0.9725 & 0.9888 & 4000 & 47.3  & 449.0  & 1.4 \\
$\rho_{T}$    & 0.0177 & 0.0006 & 0.0602 & 4000 & 100.0 & 3377.3 & 0.8 \\
\bottomrule
\end{tabular}
\end{table}

Posterior summaries for the CARAR(2) model are presented in Table~\ref{tab:carar2_diagnostics}. Similar convergence behavior was observed for this model, with all parameters exhibiting adequate effective sample sizes and acceptable Geweke statistics. As in the CARAR(1) specification, the spatial dependence parameter remained close to one, while the temporal dependence was represented through two autoregressive coefficients.

\begin{table}[!htbp]
\centering
\caption{Posterior summaries and MCMC diagnostics for the second-order autoregressive Rushworth et al. \cite{rushworth2014spatio} model.}
\label{tab:carar2_diagnostics}
\begin{tabular}{lrrrrrrr}
\toprule
Parameter & Mean & 2.5\% & 97.5\% & $n_{\mathrm{sample}}$ & Acceptance (\%) & ESS & Geweke \\
\midrule
$\nu^{2}$      & 0.1242 & 0.1161 & 0.1322 & 4000 & 100.0 & 968.0 & 0.2 \\
$\tau^{2}$     & 0.0662 & 0.0520 & 0.0821 & 4000 & 100.0 & 438.8 & 0.7 \\
$\rho_{S}$     & 0.9768 & 0.9642 & 0.9858 & 4000 & 39.4 & 466.9 & -0.7 \\
$\rho_{T,1}$   & -0.1478 & -0.2665 & -0.0414 & 4000 & 100.0 & 413.9 & 2.0 \\
$\rho_{T,2}$   & -0.4218 & -0.5256 & -0.3167 & 4000 & 100.0 & 528.7 & 0.6 \\
\bottomrule
\end{tabular}
\end{table}

\section{Implementation and results of the tailor-made MCMC fit of the proposed model}

The proposed hierarchical model was fitted using the tailor-made MCMC algorithm described in Appendix~\ref{APp.tailor-made}. Inference was based on 5,000 MCMC iterations, with the first 1,000 iterations discarded as burn-in. To reduce autocorrelation, every fifth post-burn-in iteration was retained, resulting in 800 posterior samples for inference. The adaptive proposal step sizes associated with the Metropolis-adjusted Langevin (MALA) updates stabilized during the burn-in period, indicating that the Robbins-Monro adaptation successfully identified appropriate proposal variances. The average acceptance rate for the MALA updates of the log-variance and log-range parameters was 57.02\%, closely matching the theoretical optimal acceptance probability for MALA. The random-walk Metropolis-Hastings updates for the log signal-to-noise ratio achieved an average acceptance rate of 20.45\%, which provided adequate exploration of the posterior distribution. \\

Convergence was assessed using trace plots, effective sample sizes, and Geweke diagnostics. Overall, the regression coefficients and signal-to-noise ratio parameters exhibited satisfactory mixing and rapid convergence. As expected, slower mixing was observed for the Gaussian-process covariance parameters, particularly the log-range and log-variance parameters. This behavior is characteristic of Gaussian-process covariance models, where variance and range parameters are often only weakly identifiable from the likelihood and therefore exhibit stronger posterior dependence. Despite this slower mixing, the diagnostic measures indicated stable posterior exploration following the burn-in period. 

%\bibliographystyle{vancouver}
%\bibliography{wileyNJD-Vancouver.bib}

%\end{document}

\end{document}